\documentclass[manuscript]{aastex}

\usepackage{subfigure,epsfig}
\usepackage{icarus}

\shorttitle{Oort Cloud Formation in Open Cluster Environments}
\shortauthors{Kaib \& Quinn}

\begin{document}

\bibliographystyle{icarus}

\title{The Formation of the Oort Cloud in Open Cluster Environments}

\author{Nathan A. Kaib and Thomas Quinn}
\affil{Astronomy Department, University of Washington,
    Seattle, WA 98195 \\
    Tel: (206) 616-4549 \\
    Fax: (206) 685-0403 \\
    E-mail: kaib@astro.washington.edu
\vspace{5 in}
\begin{flushleft}
65 pages\\
14 figures\\
3 tables
\end{flushleft}
\newpage
\begin{flushleft}
Running head: Oort Cloud Formation in Open Cluster Environments
\newline
\newline
Send correspondence to:\\
Nathan Kaib\\
Astronomy Department\\
University of Washington\\
Box 351580, U.W.\\
Seattle, WA 98195-1580
\vspace{5 in}
\end{flushleft}}

\begin{abstract}
We study the influence of an open cluster environment on the formation and current structure of the Oort cloud.  To do this, we have run 19 different simulations of the formation of the Oort Cloud for 4.5 Gyrs.  In each simulation, the solar system spends its first 100 Myrs in a different open cluster environment before transitioning to its current field environment.  We find that, compared to forming in the field environment, the inner Oort Cloud is preferentially loaded with comets while the Sun resides in the open cluster and that most of this material remains locked in the interior of the cloud for the next 4.4 Gyrs.  In addition, the outer Oort Cloud trapping efficiencies we observe in our simulations are lower than previous formation models by about a factor of 2, possibly implying an even more massive early planetesimal disk.  Furthermore, some of our simulations reproduce the orbits of observed extended scattered disk objects, which may serve as an observational constraint on the Sun's early environment.  Depending on the particular open cluster environment, the properties of the inner Oort Cloud and extended scattered disk can vary widely.  On the other hand, the outer portions of the Oort Cloud in each of our simulations are all similar.  
\end{abstract}

\keywords{Oort Cloud,long-period comets,Solar System formation}

\section{Introduction}

The 10$^{11-12}$ icy bodies comprising the Oort Cloud constitutes the most remote part of our solar system.  This structure's existence was first postulated to explain the steady appearance of new long-period comets in the inner solar system.  The orbits of these bodies typically have aphelia taking them tens of thousands of AU from the Sun, and their orbital inclinations are isotropically distributed.  \citet{oort50} first proposed that these bodies are members of a large spherical reservior of objects surrounding the Sun.  As impulses from passing stars alter the orbits in this reservoir, a small fraction of these bodies are injected into the inner solar system, appearing as long-period comets to observers.

\citet{oort50} put forth a scenario to explain the formation of this comet cloud as well.  Because densities in the solar nebula thousands of AU from the Sun would have been far too low to form solid bodies, these objects could not have formed in situ.  Instead, these bodies were originally leftover from planet formation in the giant planet region. As they orbited the Sun, they inevitably suffered close encounters with the giant planets, which altered their orbital energies (i.e. semimajor axes) while perihelia remained relatively constant.  As these orbits underwent random walks in energy space, some were thrown into the Sun, while others were sent to larger and larger aphelia with each passage through the planetary region.  

Left unchecked, this process would have eventually ejected all of these bodies to interstellar space.  At distances larger than $\sim$10$^{4}$ AU, however, the binding force of the Sun became comparable to the strength of perturbations from passing stars.  These perturbations applied torques to the orbits at these distances, which shifted their perihelia.  This new random walk in perihelion led to a collision with the Sun or to an orbit that no longer passed through the planetary region.  For those orbits whose perihelia were outside the planetary region, the random walk in energy space ceased since close encounters with planets no longer occurred. These orbits would then be stable unless another later outside perturbation pushed their perihelia back into the planetary region, creating a long-period comet.  The objects in these stable orbits make up the Oort Cloud.

This formation scenario of the Oort Cloud has been verified by a number of different numerical studies \citep{weiss79,fern80}. \citet{dun87} completed numerical simulations of the Oort Cloud that also included perturbations due to the galactic tide.  Their work verified that the tidal perturbations of the galaxy are actually about twice as powerful as those due to individual stellar encounters, as first postulated by \citep{heistre86}. More recently, \citet{dones04} ran formation simulations similar to \citet{dun87} but eliminated a number of numerical shortcuts contained in the previous work that were necessary due to the limited computing resources of the time.  Their results indicated that the Oort Cloud trapping efficiency was lower than previously thought ($\sim$5\%) and that many bodies migrated in toward Jupiter before they were scattered into the Oort Cloud.

The perturbations included in the previous works mimicked those of the Sun's current galactic environment.  We know, however, that most stars form in cluster environments \citep{walt00,throop01}, and the perturbations experienced in a star-forming region would be much stronger than those in a field environment.  \citet{fern97} demonstrated that a dense star-forming environment would result in a much more compact comet cloud, as external perturbations would be comparable to the Sun's binding force at much smaller distances.  This is primarily due to the more powerful stellar encounters in a cluster environment and the tidal forces of a molecular cloud.  

\citet{fernbrun00} followed up this work with numerical simulations. These simulations included perturbations from the tide of a placental gas cloud surrounding the solar system as well as encounters from a cluster star population.  The formation of the Oort Cloud was modeled in three different cluster environments of varying densities. \citet{fernbrun00} found that the structure of the resulting comet cloud depended on the stellar density of the cluster and that both the inner and outer boundaries of the cloud shifted sunward with increasing stellar densities.

\citet{bras06} completed numerical simulations of Oort Cloud formation in an embedded cluster environment, which is the star formation environment of most stars \citep{lala03}.  Unlike the open clusters used in \citet{fernbrun00}, these embedded clusters disperse on the same timescales as their gas, 1-5 Myrs.  The gravitational effects from stars as well as the gas are included in these simulations.  Run for 3 Myrs, \citet{bras06} find that the perturbations of an embedded cluster easily produce Oort Cloud objects with semimajor axes of hundreds of AU similar to that of Sedna \citep{brown04} and 2000 CR$_{105}$ giving support to the idea that these objects are genuine inner Oort Cloud bodies.  

However, all the previous numerical works that have studied Oort Cloud formation in a star-forming environment have ended their simulations shortly after the Sun exits this setting.  This is excellent for examining the qualitative effects of a dense environment on Oort Cloud formation, but estimating the quantitative effects on the Oort Cloud population from this work is difficult.  The observed sample of long-period comets is the only population constraint on the real Oort Cloud, and these bodies all originate in the outer Oort Cloud ($a > 20000$ AU).  Additionally, our observations are probing an Oort Cloud that has undergone 4.5 Gyrs of evolution rather than the shorter timespans modeled in previous simulations.  To use simulations to make population estimates, therefore, requires fully forming the outer Oort Cloud by evolving it to the current epoch and then comparing the number of bodies residing in different parts of the cloud with those in the outer cloud.

To address this, we run four different large (16000 particles) simulations of Oort Cloud formation, each with a different primordial environment that then transitions to a field environment for 4.4 Gyrs.  In addition, we have also run 15 other smaller-particle-number (2000) simulations to probe the stochasticity of this process.  For our primordial environments, we choose to build on the work of \citet{fernbrun00} that modeled Oort Cloud formation in open clusters.  In this environment, stellar encounters dominate the external perturbations, making it a straightforward setting to model.  By varying the cluster density, we are probing a wide range of perturbation strengths, the effects of which can be applied to evaluate many primordial scenarios.

Statistically speaking, it is quite unlikely that the Sun resided in an open cluster.  \citet{lada04} finds that only between 4 and 7\% of stars originate in bound clusters.  A somewhat higher, but still small fraction was calculated in \citet{lamgie06}, who find that stars form in embedded clusters at a rate that is only two to three times the rate that stars form into bound, or open, clusters.  Regardless of which figure is more accurate, it is clear that embedded clusters are the most common star formation environment and that residence in an open cluster is a relatively rare occurence.  The possibility that the Sun resided in an open cluster should still be considered, however.  Evidence from short-lived radionuclides indicates that the Sun's circumstellar disk was polluted by a nearby supernova early in its history \citep{ouell05}.  This indicates that the Sun's initial cluster must have contained at least one star massive enough to go supernova.  Since these stars are relatively rare, this implies that the Sun's original cluster was most likely very rich.  While it's unclear which proto-clusters are progenitors of open clusters, it has been suggested that the most massive embedded clusters result in open clusters which last for tens to hundreds of Myrs \citep{lada04}.  

Our current work is described in the following sections.  Section 2 focuses on the numerical techniques implemented to run our simulations, including how they differ from previous work and a discussion of their errors.  In Section 3 we examine the bulk properties of each of the Oort Clouds we form including radial distribution of material, orbital isotropization, and cloud trapping efficiency.  In addition, we measure the degree that the inner Oort Cloud ($a < 20000$ AU) diffuses to the outer Oort Cloud ($a > 20000$ AU), and we also probe the connection between our work and the observed extended scattered disk.  Finally, we summarize our work in Section 4.

\section{Numerical Methods}


To model the formation of the Oort Cloud, we use the SWIFT RMVS3 solar system integrator package \citep{levdun94} and represent comets with massless test particles.  This package is unique in its ability to efficiently and accurately integrate highly eccentric orbits as well as close encounters between planets and massless particles.  Given that bodies are placed in the Oort Cloud via planetary scatterings and that many Oort Cloud orbits are nearly parabolic, this integration package is very well-suited for this work.  Comets are removed from simulations if they are further than 200000 AU from the Sun and unbound or if they undergo a collision with the Sun or a planet.

The representation of Oort Cloud comets with massless, non-interacting test particles is well-justified, since typical comet masses are about one-billionth of an Earth-mass and individual bodies are separated by over 1 AU on average in the Oort Cloud.  For pre-Oort Cloud comets still passing through the planetary disk, however, the validity of this assumption is less clear.  Both \citep{stern01} and \citep{gold04} have used analytical treatments to show that most comets may be collisionally eroded before they ever attain eccentricities high enough to reach the Oort Cloud.  \citep{charnmorb03}, however, have developed a hybrid numerical code to model this collisional process and have reached the opposite conclusion.  We will use our own simulations to address this issue in a companion paper and attempt to quantify how important collisional processes are in these simulations.

\subsection{Initial Conditions}

Each simulation we have run includes the Sun with the four giant planets on their present orbits.  Although the orbits of the giant planets have certainly evolved thoughout the solar system's history \citep{fernip84,mal99,gome04,gom05}, we have not included their migration in order to minimize the number of different parameters we vary.  In addition to the planets, we have placed test particles from 3.25 to 40 AU.  The initial eccentricities of the test particles are uniformly distributed between 0 and 0.01, while cosine of the inclinations is uniformly distributed between 0 and 0.02.  We have run four large simulations containing 16000 particles each as well as an additional 15 simulations containing 2000 particles each.  Note that the eccentricities and inclinations of the giant planets evolve due to their mutual interactions.

\subsection{External Perturbations}

Our simulations include two types of stellar encounters: those associated with an early open cluster environment and those due to passing field stars.  In many previous studies \citep{dun87, fern80, weiss79}, the impulse approximation was employed to model the effects of stellar passages on Oort Cloud orbits.  This approximation is valid whenever the stellar velocity is much larger than the orbital speed of comets, as is the case for most passing field stars.  Observations of open clusters, however, indicate their stars have much lower velocity dispersions ($\sim$1 km/s) than field stars. With these much lower encounter velocities, the classic impulse approximation breaks down.  

To model encounters with the cluster star population, we use the method outlined in \citet{brunfern96} and \citet{hen72}.  At a distance approximately equal to one half of the average stellar spacing of the cluster, we start stars at random positions and velocities to model an isotropic velocity distribution.  The initial speeds of these stars are all set to 1 km/s, as this is the dispersion derived for a virialized open cluster in \citet{binntre87}.  Additionally, stellar masses are randomly assigned from an initial mass function (IMF) based on \citet{scalmil79} with a low-mass cutoff of 0.05 M$_{\Sun}$.  The Scalo-Miller IMF uses the form equivalent to a log-normal shown below:
\begin{equation}
\log{_{10}f\left(\log_{10}{m}\right)}= a_{0}-a_{1}\log{_{10}m}-a_{2}\left(\log{_{10}m}\right)^2
\end{equation}
where $a_0$ = 1.53, $a_1$ = 0.96, and $a_2$ = 0.47.  This IMF is convenient because it is a simple analytical form, but it has been pointed out that log-normal distributions do a poor job of fitting the real IMF above 1 $\rm M_{\sun}$ \citep{reidhawl}.  Given that the bulk of our stellar encounters involve stars with masses $< 1 \rm M_{\sun}$, however, we believe that this does not have a large effect on our results.  Furthermore, new analysis by Bochanski et al. (in preparation) of the largest stellar photometric survey to date indicates a log-normal fit is appropriate for low-mass stars.

After choosing stellar masses, the paths of these stars are then integrated until their distance from the Sun once again exceeds their original starting distance.  We assume the cluster has a lifetime of 100 Myrs, and to model its dispersal, we decrease the stellar density of the cluster linearly from the start of the simulation until it reaches zero at 100 Myrs. For our 16000-particle simulations,we have chosen to model three different initial cluster densities of 10, 30, and 100 stars/pc$^{3}$ in line with the work of \citet{fernbrun00}.  In addition, we run one 16000-particle simulation without a cluster environment as a control case.  Lastly, we also run a suite of 2000-particle simulations starting in cluster densities ranging from 0 to 100 stars/pc$^{3}$.

The other population of passing stars are those in the galactic field. To model these encounters, we use the same method employed for cluster stars.  For this stellar population, however, our critical distance at which passages are started is fixed at 200000 AU.  This distance was chosen because it is computationally inexpensive since the total number of passing stars being integrated at any one time is usually one.  In addition, most previous models of the Oort Cloud show that the vast majority of its comets orbit well inside this distance.  To assign velocities to the passing stars we relied on stellar statistics of the solar neighborhood gathered from Hipparcos observations \citep{mig00}.  The assigned stellar velocity was drawn from a Gaussian distribution dependent on stellar mass (again stellar masses were assigned using the MF in \citet{scalmil79}).  The dispersions chosen for each stellar mass bin were based on the data given in Table 6 of \citet{mig00} and are again listed in our Table 1 below.  In addition, we added extra random velocity terms to account for peculiar motion of the Sun relative to the rest of the thin disk population.  Based on Table 3 of \citet{mig00}, these were drawn from a uniform distribution ranging between 0 and the Sun's current pecular velocity values ($u_{\Sun} \simeq 10.5$ km/s, $v_{\Sun} \simeq 11.0$ km/s, $w_{\Sun} \simeq 7.4$ km/s).  For each stellar passage, all of our velocity terms are drawn and then added together.  A random starting trajectory is then assigned.  Even though the velocity distribution is clearly non-isotropic, this approach implicitly assumes that there is no directional bias in the directions of the stars' paths.  Although this is a very simplistic approach, galactic simulations suggest that the Sun's orbit may have significantly evolved over the history of the solar system \citep{wiel96,selbin02,rosk08}, and we are ignorant concerning the details of this evolution.

\begin{table}[htp]
\centering
\begin{tabular}{c c c c}
\hline
M$_*$ & $\left<u^2\right>^{1/2}$ & $\left<v^2\right>^{1/2}$ & $\left<w^2\right>^{1/2}$ \\ 
(M$_{\Sun}$) & (km/s) & (km/s) & (km/s) \\[0.5ex]
\hline
$>$ 2 & 17.0 & 11.3 & 7.2 \\
1.6 - 2 & 19.8 & 13.4 & 8.1  \\
1.4 - 1.6 & 22.5 & 15.3 & 9.9 \\
0.67 - 1.4 & 28.0 & 18.0 & 14.0 \\
$<$ 0.67 & 32.2 & 23.1 & 18.1 \\
\hline
\end{tabular}
\caption{Velocity dispersions chosen for simulated field star passages.}
\label{table:1}
\end{table}

In addition to stellar perturbations, we also model the tidal perturbations of the open cluster potential.  To choose a suitable potential we follow the work of \citet{giel06}, who modeled the dispersal of open clusters and used a plummer model to approximate the cluster stellar distribution.  In line with this work, we choose a plummer radius of 1.17 pc, which sets the virial radius at 2 pc.  Furthermore, we adjust the cluster mass so that the average density within the half-mass radius equals the desired initial cluster density for that particular run.  To account for the slow dissolution of the open cluster, we linearly decrease the cluster mass to zero over the first 100 Myrs of the simulation while holding the plummer radius constant.  We also must choose a suitable orbit for the Sun in the cluster potential.  To do this, we follow the prescription outlined in \citet{bras06} who set the Sun on random orbits in embedded cluster plummer potentials.  The cluster tidal force is then given by the difference in the cluster potential accelerations between the comet and the Sun.

Tidal perturbations from the open cluster were added to our simulations as an afterthought, and we realize that our separate treatment of the cluster stellar  and tidal perturbations is not self-consistent for a number of reasons.  First, as the cluster steadily loses mass, the velocity dispersion will certainly change, and hence the velocity of typical stellar encounters will also change.  In our code, however, the approach velocity of stars at infinity is fixed at 1 km/s and does not reflect this change.  Secondly, assuming a linear decrease in cluster mass with a fixed plummer radius is a gross simplification of the open cluster disruption process.  In reality, both of these parameters will be varying with time and a smooth linear decrease in the magnitude of cluster tides is unlikely.  We believe, however, that both of the above effects will be second-order in nature.  Setting the plummer radius to 1.17 AU and varying the density between 10 and 100 stars/pc$^3$, we find that the cluster velocity dispersion varies only between $\sim$0.5 - 1.5 km/s, thus the range in stellar approach velocity is much smaller than the ranges of stellar masses and impact parameters.  Furthermore, we will show later that perturbations from cluster stars dominate over the effects of the cluster tides, so our results are not sensitive to our exact treatment of changes in the tidal field.

The other major perturber of the Oort Cloud is the galactic tide.  To account for this pertubation, we use the tidal force formulation employed in \citet{heistre86,wietre99,lev01}:
\begin{equation}
F_{tide}=(A-B)(3A+B)\tilde{x}\bf{\hat{\tilde{x}}}\rm{-(A-B)^{2}\tilde{y}}\bf{\hat{\tilde{y}}}\rm{-(4\pi G\rho_{0}-2(B^{2}-A^{2}))\tilde{z}}\bf{\hat{\tilde{z}}}
\end{equation}
This formulation includes a smaller but significant radial term in addition to the azimuthal one. This tidal model is based on measured values of the Oort constants, which we set as $A=11.3$ km/s/kpc and $B=13.9$ km/s/kpc \citep{ollmerr98}. Additionally we set $\rho_{0}$ equal to 0.1 M$_{\Sun}$/pc$^3$.  The plane of the solar system is inclined to the galactic plane by 60.2 degrees for the entire simulation.

\subsection{Hierarchical Timesteps}

Numerically modeling the process of Oort Cloud formation is extremely CPU intensive.  This is mostly due to the disparity between the dynamical timescales of orbits in the inner planetary region and those in the actual Oort Cloud.  While the orbital periods of Oort Cloud comets range from thousands to millions of years, the period of Jupiter is only 11.9 years.  Because gravitational interactions between giant planets and comets occur throughout the entire simulation, the integration timestep must be held at a small value necessary for the accurate integration of Jupiter's orbit rather than the much larger step suitable for most cometary bodies.  

However, strategies have been devised to minimize this bottleneck. \citet{dun87} forced the giant planets to follow non-interacting circular orbits.  These predetermined orbits allowed them to use larger time steps for cometary orbits without actually integrating the giant planets.  Given that Oort Cloud comets re-enter the solar system at essentially random times with repect to periods, the small eccentricities and inclinations of the real planets can be neglected in these occurances.  In addition, \citet{dun87} represented the planets and Sun as a point mass in a barycentric frame whenever a comet is outside of 150 AU.  This is an acceptable approximation, as the quadrupole gravitational term is only $\sim$10$^{-6}$ of the monopole term at this distance \citep{dun87}.

On the other hand, the assumption of circular planetary orbits is not valid when modeling the initial scattering of material out of the planetary region during the beginning of our formation simulations.  During this period of frequent, multiple planetary perurbations of comets' orbits, the non-zero eccentricities and inclinations of the planets have a significant effect on the scattering process.  The effects of Jupiter's eccentricity on scattering processes in the terrestrial zone have been well-documented \citep{chamweth01,levagn03,ray04}, and our preliminary runs showed that the scattering caused by circularly orbiting planets is significantly different than when the giant planets are placed on their current orbits.

We have developed an efficient algorithm that both minimizes the timescale disparity bottleneck and still accurately models the scattering of bodies out of the planetary disk.  We have modified SWIFT to split our test particles into two subgroups every $\sim$10 years, those coming within 300 AU of the Sun and those staying outside 300 AU.  For particles  entering or already  within 300 AU, the code integrates them and the giant planets in heliocentric coordinates at a time step of 100 days.  It repeats this integration for 36 steps for a total time of $\sim$10 years.  The planets' positions and velocites just before and after this series of 36 integrations are stored.  Using the saved positions and velocities of the planets, the other population of particles outside 300 AU is then integrated with a large 10-year time step in the barycentric frame \citep{wietre99}.  Our use of a smaller time step near the Sun for large orbits is somewhat analogous to the planetary close encounter routine implemented in the unmodified RMVS3 routines \citep{levdun94}, and in this way it can be thought of as a ``solar system'' close encounter routine.  

Care must be taken to try to insure the time-reversibility of the integrator, .i.e. the same sequence of timesteps must be chosen if we were to run the particles in reverse.  To do this, we first drift all particles by a large timestep.  Any that have penetrated the critical radius at the end of this drift are then set back to their original positions and integrated instead with a small time step.  This prevents particles integrated on a large step from having a portion of their orbit near the planets integrated with a large timestep that does not properly account for planetary perturbations.  Failure to implement a time-reversible step selection scheme will indeed lead to large unphysical secular drifts in the orbital elements (Quinn et al. 1997, astro-ph/9710043).  

Although time-reversibility is nearly always preserved in this algorithm, there are rare instances when time symmetry is broken.  These instances occur when an incoming comet is integrated with a large timestep and calculated to be just outside 300 AU after the integration.  Because the orbital integration will be slightly different with the small 100 day timestep, it is possible that the same integration using small steps would leave the same comet just inside 300 AU.  In these occurrences, our code uses a large timestep when the small one should have been employed.  Although this scheme is not perfectly time reversible, it nevertheless offers significant advantages over more simplistic routines  (Quinn et al. 1997, astro-ph/9710043).  Furthermore, it lacks the symplectic properties of a truly reversible multiple timestep scheme such as that employed in \citet{dunetal98}.

In addition to time-reversibility, we have also taken care to still include planetary perturbations when we integrate in the barycentric frame just as in the heliocentric, albeit sampled at a lower rate.  In the RMVS3 routine we drift test particles about a point mass that contains the mass of the solar system.  To still attempt to account for the distant slight planetary perturbations, we use the following equation in the kick routine to calculate this force, which is just the difference between a monopole field centered at the solar system's barycenter and the real field,

\begin{equation}
F_p=M_{SS}\frac{\bf{r}}{r^3}-M_S\frac{\bf{r}-\bf{r_S}}{\left|r-r_S\right|^3}-\displaystyle\sum_{p}M_p\frac{\bf{r}-\bf{r_p}}{\left|r-r_p\right|^3}
\end{equation}
where the $S$, $p$, and $SS$ subscripts correspond to the Sun, planets and the solar system barycenter.  This allows the potential to still deviate from a monopole slightly as the planets and Sun orbit about one another, just as in the heliocentric frame.  For distant orbits, the barycentric frame is preferred because the kick acceleration is not dominated by a large rapidly oscillating indirect term as in the heliocentric frame.

This modification to SWIFT increases the computational efficiency of outer integrations by a factor of 36.  Separating the comets into subgroups slows the code some, but the fraction of material in the planetary region decreases to $\sim$25\% within the first 100 Myrs. Because most particles do not closely approach the Sun after the first 100 Myrs, this modified version of SWIFT quickly overtakes a version that uses a single time step.  Overall, this change offers a factor of 10 increase in computing efficiency.  Additionally, the modified version of SWIFT produces comet clouds that are essentially identical to those produced by the original version, thus the dynamics of Oort Cloud formation are unchanged by our modification.

\subsection{Integrator Inaccuracies}

Even with the preservation of time-reversibility, our heirarchical timestepping routine does introduce noticeable errors into integrations of orbits that continually cross the 300 AU timestepping boundary.  Scattered disk bodies whose aphelia are just beyond 300 AU are most sensitve to this error because they will be making the most timestep transitions due to their small orbital periods.  To characterize this error, we have integrated 200 test particles on hypothetical scattered disk orbits for 1 Gyr.  The initial perihelia were uniformly distributed between 40 and 120 AU, while initial semimajor axes ranged between 300 and 1000 AU.  We have performed these integrations twice; once with an unmodified version of SWIFT and again with our version containing a variable timestep.  We have measured the integrator's inaccuracies by examining the perihelion change of the orbits over time.  Because there are no external perturbations and the particles do not pass very near the planets, the perihelion should stay nearly constant for the entire integration.  The exception to this occurs near our inner boundary of 40 AU where Neptune's orbital influence can cause perihelia to chaotically diffuse over long timescales.  The final perihelia of the orbits after 1 Gyr are compared in the top panel of Figure 1.  As can be seen in this figure, the errors introduced by the timestepping produce a small random walk in perihelia.  We say random because there appears to be no bias toward positive or negative perihelion drift.  Furthermore, when fitting a power-law to the growth of the rms perihelion difference over time, we find that this error is proportional to $t^{0.5}$.  This is shown in the bottom panel of Figure 1.  After 1 Gyr of integration, the rms difference in perihelia between the original and modified integrators is 0.4 AU.  Extrapolating this out leads to an error of $\sim0.8$ AU after 4 Gyrs.  In Figure 1b, we also plot the difference between initial perihelion and final perihelion for the particles integrated with the unmodified version of SWIFT.  This allows us to compare our artificial drift values with those caused by real secular perturbations.  As can be seen, the artificial changes are roughly a factor of 2-3 times the magnitude of the secular changes in perihelion inside $\sim$50 AU.  We have also integrated the orbit of a real object, Sedna, with both integrators and find a difference of $\sim$0.1 AU between the two perihelia after 1 Gyr.

\begin{figure}[htbp]
\centering
\includegraphics{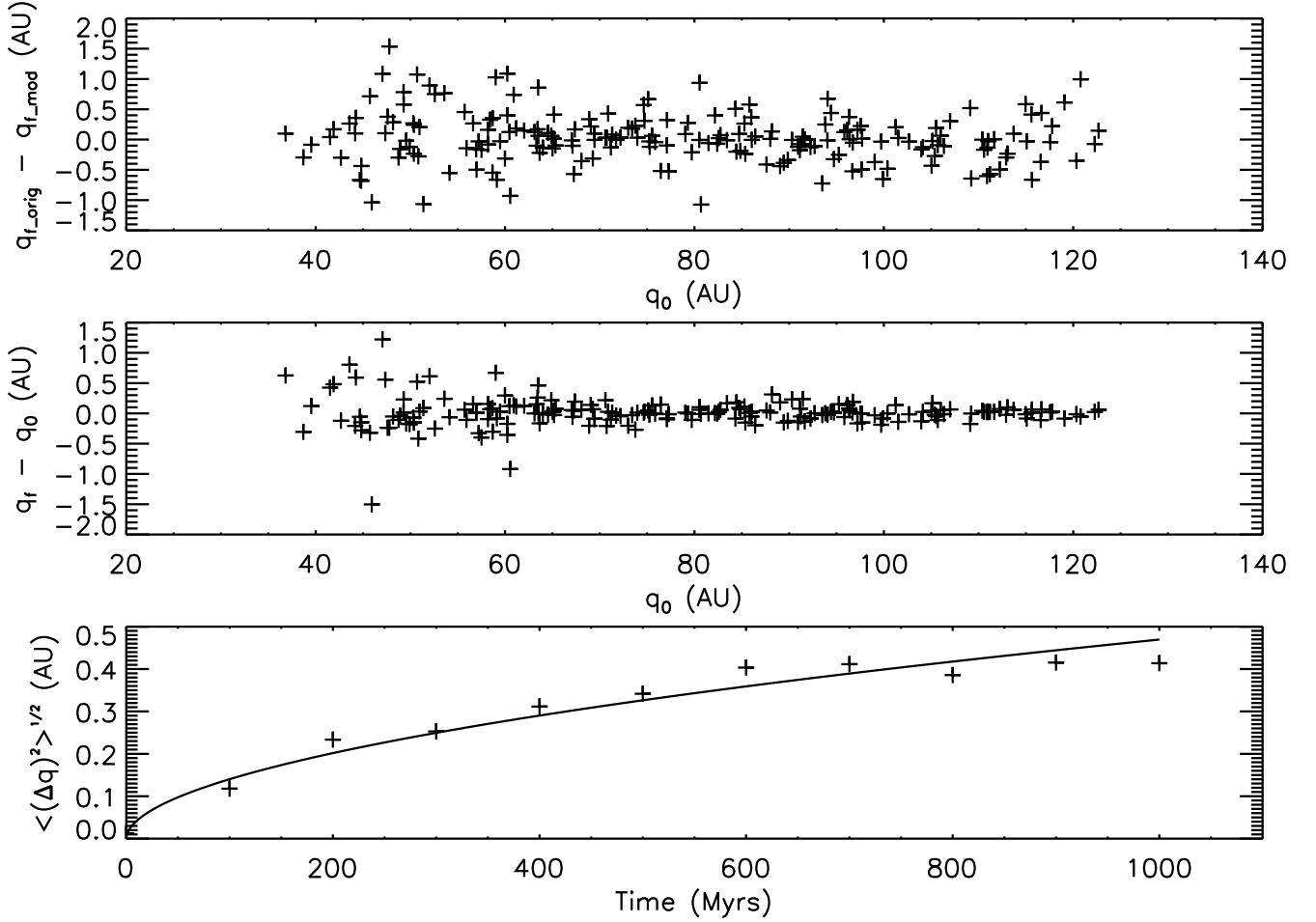}
\caption{\it{Top: }\rm The difference in final perihelion between particles integrated with an unmodified version of SWIFT RMVS3 and the same particles integrated with our modified code.  Particles are integrated for 1 Gyr.\it{Middle:}\rm The difference between initial and final perihelion vs. initial perihelion for the particles integrated with an unmodified verion of SWIFT RMVS3. \it{Bottom:}\rm The growth in the rms difference in perihelia for our test particles when comparing the modified version of SWIFT with the original.  The solid line is a best-fit power-law to the data, $<(\Delta q)^2>^{1/2} \propto t^{0.52}$.}\label{fig:1}
\end{figure}

A perihelion migration of this magnitude is inconsequential for perihelia outside ~45 AU since they are unable to migrate significantly toward Neptune.  For objects in the traditional scattered disk ($q \lesssim 40$ AU), however, a migration of this scale could pose a problem over the lifetime of the solar system.  This could cause scattered disk objects just outside the edge of Neptune's influence to drift inward and fall under the control of the planet.  This would eventually lead to ejection or placement in the Oort Cloud, and would artificially decrease the number scattered disk objects as well as alter the distribution of perihelia in the scattered disk.  

Fortunately, the orbits that are most sensitive to this effect are typically not dynamically stable long enough for these errors to substantially accumulate.  Perihelia only begin to artificially drift when an object has reached an aphelion beyond 300 AU.  Objects with perihelia inside 40 AU and aphelia beyond 300 AU have already undergone significant scattering events from Neptune and will not remain stable in these types of orbits for a long enough time for integration errors to become significant.  Before the errors accumulate to high levels, these bodies will be ejected or placed in the Oort Cloud or ejection.  Basically, by the time an object reaches $Q > 300$ AU, it is already well on its way to the Oort Cloud.  We have looked at all of the particles in each of our simulations that have spent any time with 30 AU $<$ $q$ $<$ 40 AU and $Q$ $>$ 300 AU.  The mean residence time in this type of orbit is only $\sim$200 Myrs in every simulation.  Because the timescale of orbit removal is significantly less than the timescale for significant perihelion drift from integrator error, we conclude that our integration inaccuracies do not affect the results of our simulations.  

The group of particles that is an exception to this are those in the inner extended scattered disk ($40 \lesssim q \lesssim 45$ AU).  In some of our simulations strong early external perturbations pull the perihelia of tightly bound orbits out to this range.  Objects are then left on orbits with semimajor axes of hundreds of AU and perihelia between 40 and 45 AU for the remainder of the simulation.  Most of these bodies will be stable over the lifetime of the solar system and will not be affected by Neptune.  With the perihelion drift caused by our code modifications, however, it is possible that the perihelia of a small fraction of these bodies could drift inwards allowing them to fall under the control of Neptune.  This would eventually lead to ejection or placement in the Oort Cloud, thereby artificially diminishing this part of the extended scattered disk over time.  

This diffusion of perihelia appears to result from the slight changes of the numerical Hamiltonian as particles cross the timestepping boundary due to both the change in step size and the switch from a heliocentric frame to a barycentric one.  This causes small changes in both semimajor axis and eccentricity that slowly accumulate over time.  One last check is to ensure that these slight energy kicks due to timestep changes are not comparable to the energy kicks from planetary scatterings.  To do this, we calculate the $\Delta$(1/$a$) after one complete orbit integration for 2000 nearly parabolic particles with 100 AU $ < q < $ 200 AU and random inclinations and arguments of perihelion.  With perihelia in this range, the planetary perturbations will be negligible, and the timestepping kick will be the main perturbation.  For the 2000 orbits, we find $\Delta\left(1/a\right)_{rms} = $1.2 x 10$^{-7}$ AU$^{-1}$, which is approximately two orders of magnitude lower than the mean energy kick due to Neptune \citep{dun87,fernbrun00,brasdun08}.  It should be noted that the implementation of a symplectic corrector such as those outlined in \citet{mikpal00} and \citet{wis06} should further reduce this timestepping error, and we intend to pursue this in the future.  Lastly, it is also important to mention that the changes in other orbital elements, such as inclination, appear to be comparable to the magnitude of secular changes due to the planets seen with the unmodified version of SWIFT.

\section{Simulation Results and Discussion}
\subsection{Radial Distribution of Comets}

We define a particle to have reached the Oort Cloud when its perihelion ($q$) exceeds 45 AU, for beyond this distance the giant planets have little effect on the evolution of orbital elements.  In the following analysis, unless specified otherwise, we only consider Oort Cloud particles as defined by this criterion.  Our simulations allow us to examine the possible effects that residence in an open cluster would have on the properties of the Oort Cloud.  We begin this examination with the radial distribution of comets throughout the Oort Cloud.  In the top panel of Figure 2 we have plotted the number density of comets as a function of mean heliocentric distance for our four large (16000-particle) simulations.  Each of these densities is normalized to the density value at 20000 AU.  It can be seen in this plot that the density distribution for comets outside of $\sim$10000 AU differs very little between the individual simulations.  This is not surprising; Because of the gentle scattering of planetesimals by Uranus and Neptune, the complete assembly of the Oort Cloud takes $\sim$1 Gyr \citep{dun87,dones04}, and these simulations spend only their first 100 Myrs in different cluster environments.  During the rest of the time, each comet cloud is exposed to perturbations characteristic of the Sun's current field environment, and it is these perturbations that dominate the dynamics of the outer Oort Cloud ($a \gtrsim$ 20000 AU).  Fitting a power-law to these outer Oort Clouds yields a comet density that is proportional to $r^{-3.4}$, which is close to the $r^{-3.5}$ distributions found in previous simulation work \citep{dun87,bras06}.

Inside 10000 AU the density distributions extend inward for different distances.  The influence of the galactic tide and field star impulses on these comets is weaker, so the early strong stellar perturbations from a cluster become more important in sculpting the inner Oort Cloud, which we define as $a <$ 20000 AU.  Once these comets reach the Oort Cloud, most remain essentially locked in small semimajor axes after the solar system leaves its cluster environment.  The density profiles of the comet clouds all begin to fall off from a power-law once the inner boundary of the Oort Cloud is reached.  This occurs at a progressively smaller radius for simulations that experience more powerful early cluster perturbations.  This is again because strong cluster perturbations are efficient at placing comets into the Oort Cloud at small semimajor axes.  In the case of the densest cluster environment, the Oort Cloud density rises all the way to within 100 AU of the Sun and does not turn over until $\sim$75 AU.  The density of material in the inner 1000 AU of this simulation exceeds all of the others by orders of magnitude.

That an open cluster environment is efficient at populating the Oort Cloud at small semimajor axes is illustrated well in the bottom panel of Figure 2.  In this panel, we show the cumulative distributions of orbital energies for our four large simulations along with the results from \citet{dun87}.  \citet{dones04} point out that the initial conditions in \citet{dun87} fix comet perihelia and cause Neptune to scatter more comets into the Oort Cloud than it would with more realistic initial comet orbits.  The gentler energy kicks of Neptune in turn produce an Oort Cloud that is more centrally concentrated.  Our results from our control simulation with no cluster environment support this conclusion, as the distribution of orbital energies in this simulation matches much more closely with that seen in Dones et al. (2004) (Dones, private communication) compared to the distribution in \citet{dun87}.

\begin{figure}[htbp]
\centering
\includegraphics{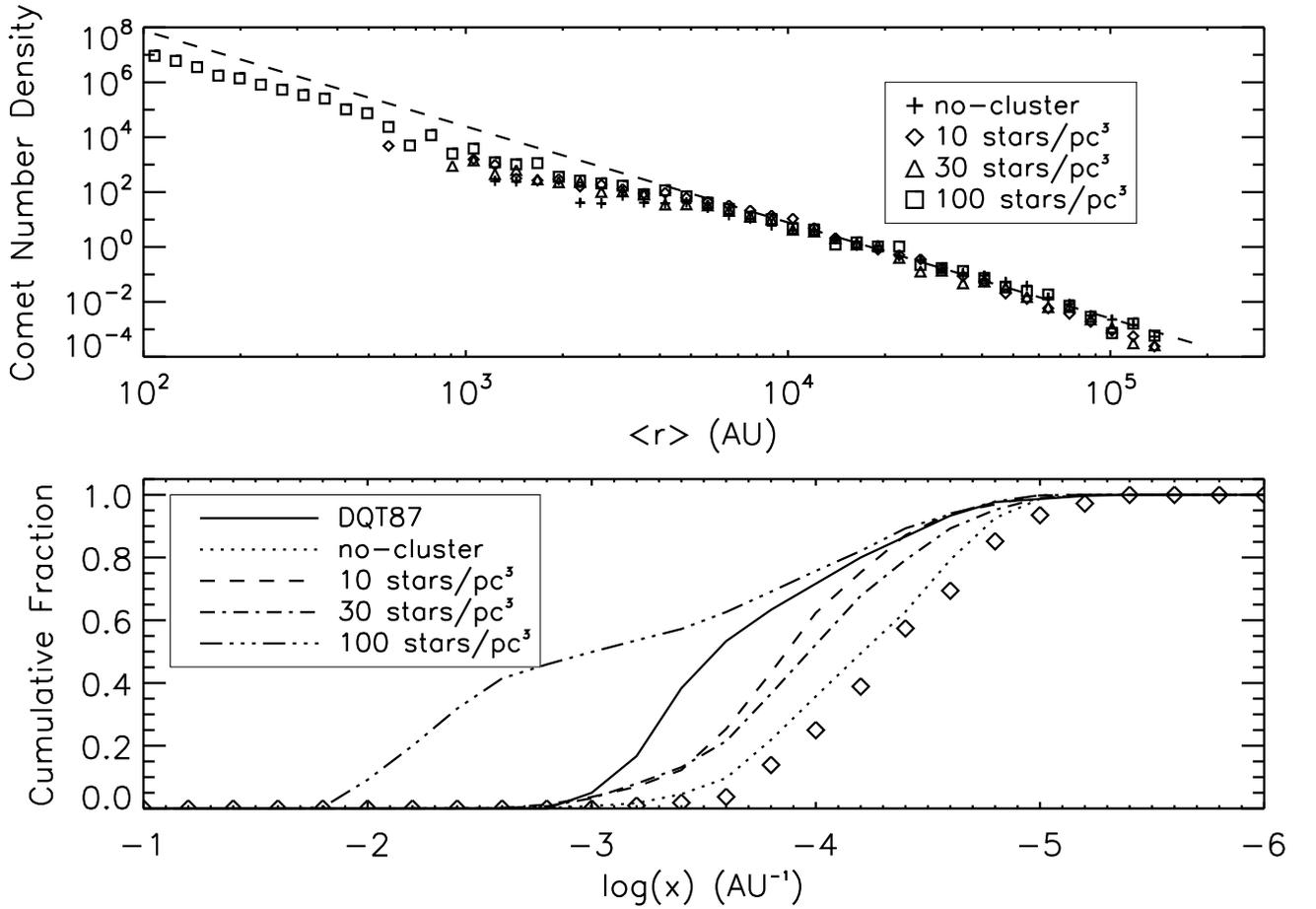}
\caption{\it{Top:} \rm{Plot of comet number density vs. mean radial distance for each of our four large simulations at $t = 4.0$ Gyrs. Each density is normalized to the density value at 20000 AU for that particular simulation.  Cluster environment densities are as follows: no cluster (crosses), 10 stars/pc$^{3}$ (diamonds), 30 stars/pc$^{3}$ (triangles), 100 stars/pc$^{3}$ (squares). The dashed line corresponds to a density distribution of $n\sim r^{-3.4}$.  This is a best fit of the combined data beyond $10^4$ AU.}  \it{Bottom:} \rm{Cumulative orbital energy distributions for each simulation at $t=4.5$ Gyrs. The results of \citet{dun87} are shown by the solid line.  The different simulation cluster environments are as follows: }\it{Dotted:} \rm{No cluster environment,} \it{Dashed:} \rm{10 stars/pc$^{3}$,} \it{Dash-dot-dash:} \rm{30 stars/pc$^{3}$,} \it{Dash-dot-dot-dot-dash:} \rm{100 stars/pc$^{3}$,} \it{Diamond data points:} \rm{Dones et al. (2004) simulation data (Dones, private communication).}}\label{fig:2}
\end{figure}

In the bottom panel of this figure, we also note two other features.  The first concerns the extreme similarities between the orbital distributions in the 10 and 30 stars/pc$^3$ simulations.  After these simulations leave their respective cluster environments, the comet cloud of the 30 stars/pc$^3$ run is noticably more compact than that of the 10 stars/pc$^3$ cluster run.  Thereafter, however, the 10 stars/pc$^3$ simulation experiences 3 field star encounters between 450 and 560 AU of the Sun.  Because of these powerful field star perturbations, an anomalously high number of comets are placed on low semimajor axis orbits after cluster dispersal.  These close field star encounters are an extemely improbable occurence.  For comparison, in our other 18 runs, only three simulations suffer a field star passage inside 600 AU, and each of these is a lone occurence.  Because of this, our large 10 stars/pc$^3$ simulation is considered an anomolous simulation not entirely representative of typical Oort Cloud formation.

The second striking feature in the bottom panel of Figure 2 is the drastic difference between the 100 stars/pc$^3$ simulation and the 10 and 30 stars/pc$^3$ runs.  The change in the orbital energy distribution in our densest cluster is huge compared to the differences in all the others shown.  Furthermore, a significant percentage of the Oort Cloud is found on orbital semimajor axes less than 100 AU.  These are smaller than the minimum semimajor axes generated in the \citet{fernbrun00} simulations of a 100 stars/pc$^3$ environment ($\gtrsim100$ AU).  As will be explained in more detail later, Oort Cloud formation inside an open cluster environment is a very stochastic process, and this result is indicative of that fact.  

It has already been determined with past simulation results that increased central concentration can be produced by residence in an open cluster environment.  Indeed, as shown in \citet{fernbrun00}, this concentration increases with cluster density.  Similarly, \citet{bras06} find that the median distance of their Oort Cloud comets scales with $\rho^{0.5}$ where $\rho$ is the density of a particular cloud's embedded cluster environment.  We look for a similar trend in our data in Figure 3 where we plot the median comet distance vs. open cluster density.  Unlike the work of \citet{bras06}, there is no well-defined correlation in our results.  Indeed, the actual median distances in each of our simulations are fairly constant near 10$^4$ AU.  This is a result of the fact that regardless of our initial cluster environment, a shell of more distant comets is added to the Oort Cloud after the Sun exits its cluster.  The range of semimajor axes in this outer shell is roughly the same for each of these simulations since they all experience similar perturbations from field stars and the galactic tide.  To isolate the effects of the cluster environments, we also plot the radius of the inner 10\% of the Oort Cloud ($r_{10}$).  There is a significantly larger spread in the values of $r_{10}$ for our simulations.  Furthermore, there is a trend of increasing values of $r_{10}$ with decreasing values of cluster density, but the tight correlation seen in \citet{bras06} is not present in our results.

\begin{figure}[htp]
\centering
\includegraphics{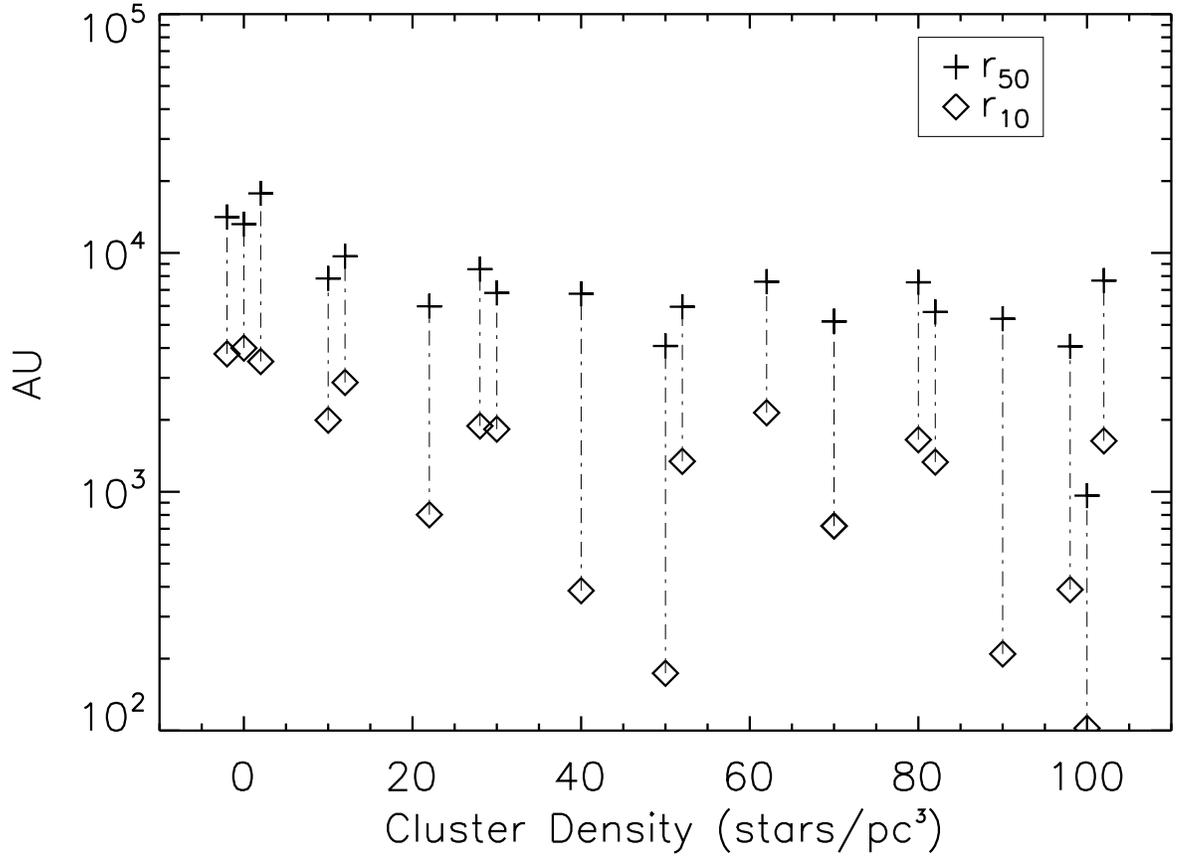}
\caption{The median Oort Cloud radius, $r_{50}$, and the radius divided the inner 10\% of the Oort Cloud from the outer 90\%, $r_{10}$, are plotted vs. the initial open cluster density of each simulation.  Data are slightly offset to avoid confusion.  $r_{50}$ and $r_{10}$ data points for the same simulation are connected with a vertical line.}\label{fig:3}
\end{figure}

The large amount of scatter in our Oort Cloud concentrations is a result of the difference between our open cluster environment and the embedded cluster environment employed in \citet{bras06}.  In \citet{bras06} the perturbations of gas tidal forces were stronger than the perturbations of stellar passages due to the high gas content of embedded clusters.  Because of this, the strongest perturbations experienced in two different simulations run with the same cluster parameters should be quite similar since the strongest perturbations are mostly determined by the overall density and distribution of gas.  On the other hand, open clusters have a very low gas content and the tidal forces experienced by the solar system come from the extended distribution of stars in the cluster.  Because of this difference, it is actually stellar passages that account for the strongest perturbing forces in our simulations.  The relative strengths of the perturbing forces are well-illustrated in Figure 4.  The distribution of semimajors axes in the Oort Cloud is shown for two different simulations at $t$ = 100 Myrs.  The first is our large simulation run in the 100 stars/pc$^3$ cluster environment.  The other is the exact same simulation only with the passing stellar masses all set to zero, so that only the cluster tides perturb the Oort Cloud.  As can be seen in this figure, for the simulation that does not contain passing stars, the inner edge of the Oort Cloud extends to barely $\sim10^3$ AU, whereas in the simulation that includes stellar passages, the inner edge of the Oort Cloud is found at $\sim100$ AU.  Thus, the tides by themselves actually begin signficantly torquing comet orbits about an order of magnitude further from the Sun than the strongest stellar passages.

\begin{figure}[htp]
\centering
\includegraphics{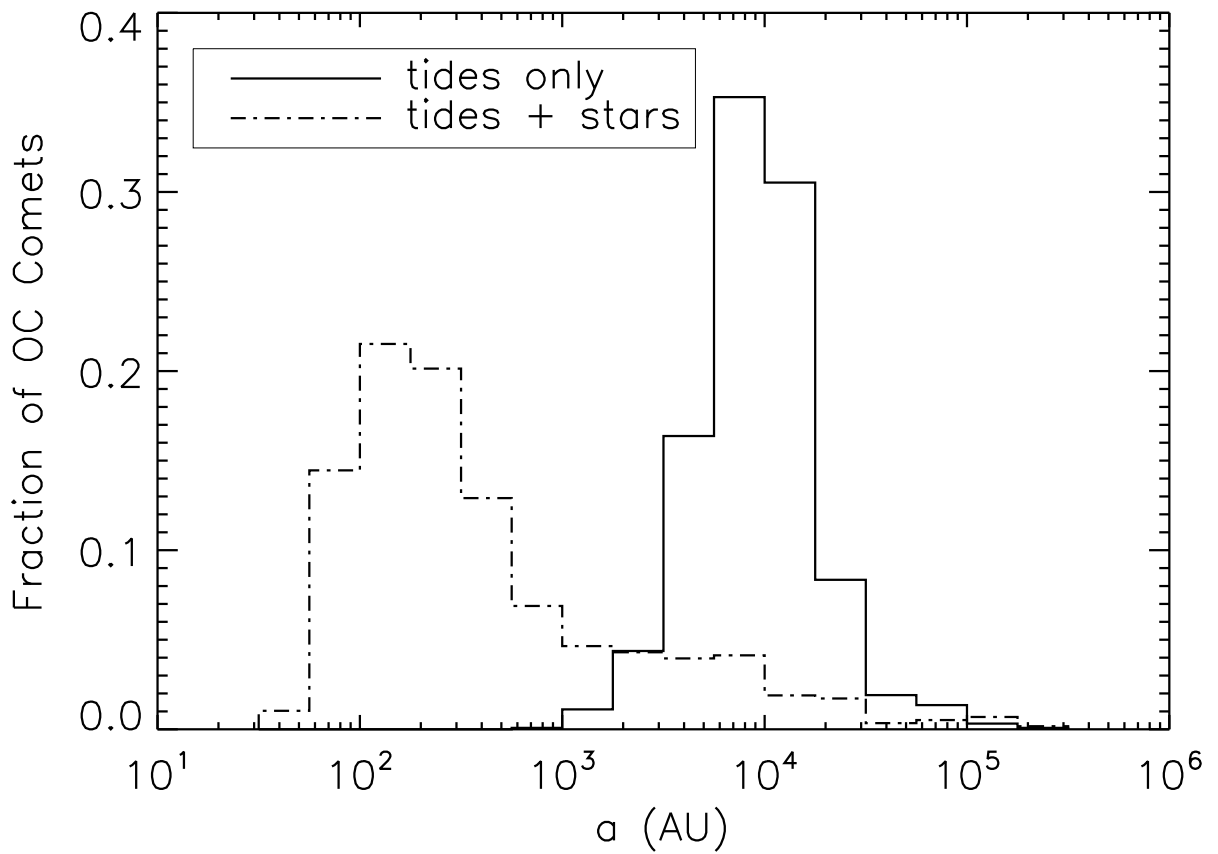}
\caption{Orbital energy distributions for the Oort Clouds of two simulations at $t=4.5$ Gyrs beginning in a 100 stars/pc$^3$ open cluster. The first simulation (solid line) contains only the cluster tidal forces and no cluster star passages, whereas the second simulation (dashed line) contains both tides and stellar passages.}\label{fig:4}
\end{figure}

In general, the closest passing star produces the strongest perturbing force.  Because of this, we posit that the closest stellar encounter will also set the inner boundary, or minimum semimajor axis, in the Oort Cloud.  This hypothesis is not easy to verify, however, because the perturbing force of cluster star encounters cannot be quantified well analytically.  In previous works studying the Oort Cloud, the impulse approximation \citep{rick76} has been employed to model the perturbations passing stars impart on comet orbits.  To be valid, however, the time duration of the stellar encounter must be much smaller than the orbital period of the comet.  In the case of our densest cluster environment, we find that comets with semimajor axes less than 100 AU are perturbed on to Oort Cloud orbits.  Given that our initial cluster density is as high as 100 stars/pc$^{3}$ and that it decreases linearly to zero over 100 Myrs, we expect the closest stellar encounter to be about 1200 AU from the Sun.  If we then use $\tau\sim(b/v_{\infty})$ as an estimate of the encounter time, this yields 7600 yrs.  In contrast, the period of a comet orbiting at 100 AU is only 10$^3$ yrs.  Thus, the impulse approximation is not very useful for our purposes.

For the situation when the orbital period of a perturbed comet is significantly smaller than the stellar encounter time, \citet{kobida01} developed an analytical approximation of the perturbations by making use of Gauss's equations \citep{brouclem61}.  To make their problem analytically tractable, however, they had to assume that the initial eccentricities of the comet orbits were zero.  In the case of Oort Cloud formation, the perturbed orbits we are interested in studying are those in the scattered disk, which typically have eccentricities closer to 1.  Although we are working in a different regime of orbital elements than \citet{kobida01}, we can compare their results to our simulations to gain insight into how non-zero orbital eccentricities are affected by stellar perturbations.  According to their work, the most powerful orbital perturbations occur for comets whose orbital mean motion is in resonance with the hyperbolic mean motion of the passing star.  In particular, \citet{kobida01} find that the cometary orbits most strongly perturbed are found near the 5:1 and lower order resonances.  In the resonant regime, the orbital perturbations are very violent, and even the semimajor axes are strongly perturbed.  To probe for resonant effects, we have plotted the minimum Oort Cloud semimajor axes against the location of various resonances with the strongest stellar encounter in each simulation.  The mean motion for a stellar passage is given by
\begin{equation}
\Omega_*=\sqrt{G\left(M_{\Sun}+M_*\right)\left(1+e_*\right)/D^3}
\end{equation}
where $M_*$ is the stellar mass, $D$ is the star's pericentre, and $e_*$ is the eccentricity of the star's hyperbolic orbit, which is
\begin{equation}
e_*=\frac{\sqrt{1+2l_*^2E_*}}{G^2\left(M_{\Sun}^2+M_*^2\right)}
\end{equation}
where $l_*$ is the star's specific angular momentum and $E_*$ is the star's specific energy.  We define the minimum semimajor axis as the mean semimajor axis of the inner 10\% of comets.  We find that the strongest correlation is with the 2:1 resonance, which is shown in Figure 5.  There is quite a bit of scatter in this correlation, and indeed the simulations of \citet{kobida01} show that other parameters such as the inclination of star's path with respect to the ecliptic will impact which resonances are most powerful.  Furthermore, the correlation is less obvious at larger semimajor axes.  Nevertheless, we believe this can serve as a useful tool to constrain the types of stellar encounters the Sun may have experienced early in its history.  

\begin{figure}[htp]
\centering
\includegraphics{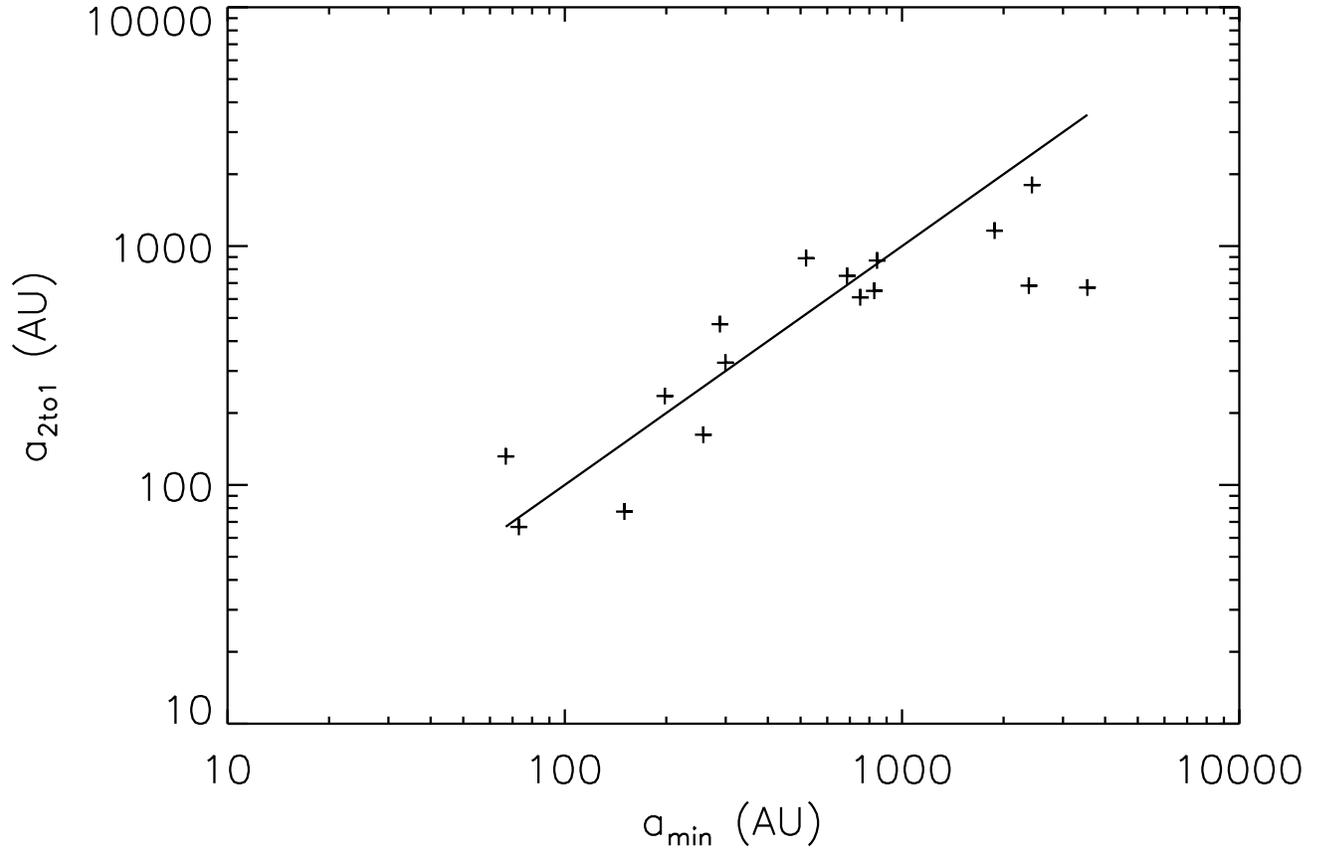}
\caption{A plot of the minimum Oort Cloud semimajor axis observed in each simulation vs. the cometary semimajor axis corresponding to the 2:1 mean motion resonance with the hyperbolic mean motion of the closest passing cluster star in a given simulation.  The solid line represents a one-to-one correlation.}\label{fig:5}
\end{figure}

That the most inner regions of the Oort Cloud are sculpted largely by the single closest stellar passage illustrates the stochasticity of forming the Oort Cloud in an open cluster environment.  For instance, we run two different simulations that have cluster densities of 50 stars/pc$^3$.  In one simulation, the closest stellar encounter occurs when a 0.17 M$_\Sun$ star passes within 1665 AU of the Sun.  In the second simulation, the strongest stellar encounter involves a 2.9 M$_\Sun$ star passing within 250 AU of the Sun.  Because of this, the minimum semimajor axis of each Oort Cloud differs by an order of magnitude.  A simple flux/cross-sectional area calculation leads us to expect that the closest encounter will be 1630 AU.  Note, however, that there is significant gravitational focusing of the stars that is not taken into account in this calculation.  While this is the closest encounter typically expected, in reality, one-fifth of simulations in this environment will suffer a stellar encounter within 1000 AU of the Sun.  Moreover, the gravitational focusing due to the star's and Sun's mass will actually make this fraction substantially higher.  This effect leads to a large amount of variance in the distributions of comets for simulations run in similar cluster environments.

The parameters for the closest stellar encounter in each simulation are listed in the 3rd  and 4th columns of Table 2 along with the minimum semimajor axis of each Oort Cloud in column 5.  

\subsection{Oort Cloud Isotropization}

In addition to examining the radial distribution of comets, we also look at how an open cluster environment can impact the isotropization of comets in our Oort Clouds.  In the top panel of Figure 6, we bin the comets by semimajor axis and then determine the mean cosine of the inclination with respect to the ecliptic for each bin.  We do this for two separate simulations: the large simulation with no cluster environment and the large simulation immersed in a 100 stars/pc$^3$ cluster.  Examining the no-cluster simulation first, we see that by 10$^4$ AU, that the mean cosine approaches zero, which indicates a mean inclination of 90$^{\circ}$ and, hence, isotropization of the Oort Cloud.  In the second case, we see that there are actually two different regions of the Oort Cloud that are isotropized separated by an intermediate region that is slightly flattened.  This is an artifact of the Sun's dynamical history in this simulation.  Between 100 and $\sim$300 AU the Oort Cloud is concentrated toward the ecliptic.  These represent comets that were pulled out into the Oort Cloud by the few strong cluster star encounters but were not perturbed strongly enough thereafter to completely randomize their inclinations.  Beyond this distance, the Oort Cloud is fairly isotropic until about 800 AU.  The comets in this inner isotropic region were also added in the cluster phase, but were far enough from the Sun that weaker cluster perturbations were powerful enough to randomize their orbits.  Moving beyond 800 AU, the comet orbits once again show a bias toward the ecliptic until about 10$^4$ AU.  In this regime a hybrid population exists.  In the low semimajor axis end are mostly comets added toward the end of the cluster phase as the cluster perturbations become steadily weaker and represent a mostly isotropized population.  Superimposed on this, however, is the low semimajor axis tail of the classical Oort Cloud, which is steadily assembled after the solar system exits its cluster environment.  Just as in the no-cluster simulation, these comet orbits are concentrated toward the ecliptic in this regime and produce a second overall flattening of the cloud.  Beyond 10$^4$ AU, the population is dominated by classical Oort Cloud comets, which, as in the no-cluster simulation, are isotropized.

This inner isotropized zone of comets exists for many of our other simulations as well.  To measure the semimajor axis where isotropy begins to occur, we bin our comets in semimajor axis as discussed above.  Then any bin with a mean $\cos{i}<$ 0.3, which corresponds to an inclination of $\sim$72$^\circ$, is said to be isotropized.  To deal with the large uncertainties due to small bin numbers in our small-particle-number simulations, we also require that the bin contain $N_{Oort}/(2N_{bins})$ comets.  In the second panel of Figure 6, we plot the isotropization semimajor axis in units of the minimum semimajor axis for each simulation as calculated for Figure 5.  We see that for most simulations within 1 to 5 minimum semimajor axes, the Oort Cloud has become isotropized.  \citet{bras06} find isotropization distances of between $\sim$20000 and $\sim$200 AU depending on their embedded cluster density.  Given that most of the minimum semimajor axes in our simulations fall between 100 and 1000 AU, this implies that Oort Clouds formed in our type of cluster environment will become isotropized between a few hundred to a few thousand AU as well.  We have essentially replicated their same range of istropization distances, although the upper limit is not realized because of the classical Oort Cloud formation that proceeds after cluster dispersal.  \citet{bras06} also observe a distinct correlation between istropization radius and cluster density.  This trend is not strong in our results and is once again the result of the stochastic nature of stellar encounter-dominated perturbations in the open cluster environment.   

In the bottom panel of Figure 6, we look at the angular momentum distribution of comets in the two large simulations plotted in the top panel.  Unlike the top panel, however, this is with respect to the galactic plane rather than the ecliptic.  In the no-cluster case, we see there is a bias toward $L_z/L=0$, or a galactic inclination of 90$^\circ$.  As discussed in \citet{dybpret96}, this is a result of the Kozai mechanism, which causes comets to spend much longer periods of time with galactic inclinations near 90$^{\circ}$ compared to the time spent with inclinations closer to 0 or 180.  Although the randomizing effects of field stars act to diminish this bias as discussed in \citet{matwhit92}, it is nevertheless still present due to the fact that galactic tidal perturbations dominate over those due to passing field stars \citep{heistre86}.  

The angular momentum distribution for the large 100 stars/pc$^3$ simulation is also shown in this panel and looks distinctly different from the classical case.  Although there still appears to be a small bias of comets toward $L_{z}/L=0$, there is a much larger spike at $L_{z}/L=0.5$ or $i_{gal}\simeq 60^{\circ}$.  Again, this is a signature of the cluster star perturbations that dominated the Oort Cloud formation process early on.  During the cluster phase, comets enter the Oort Cloud along the scattered disk, which is concentrated in the ecliptic and therefore has a galactic inclination of 60.2$^\circ$.  Rather than being smoothly pushed toward $L_z/L = 0$ by the galactic tide, however, these comets are given random impulses resulting from stellar encounters and have no directional bias in angular momentum.  This results in essentially a random walk away from a galactic inclination near $60^\circ$.  Once the solar system then exits its cluster environment, most of these comets are close enough to the Sun to be insulated from further orbital alteration by the galactic tide, thus the signature of this diffusion from the ecliptic is preserved.

A summary of our isotropization results can be found in the last colum of Table 2.  

\begin{table}[htp]
\centering
\begin{tabular}{c c c c c c c c c c}
\hline
Sim. & $N$ & $n_*$ & $D_{*min}$ & $M_{*}$ & $a_{min}$ & $a_{iso}$\\ 
 & & (pc$^{-3}$) & (AU) & (M$_{\Sun}$) & (AU) & (AU)\\[0.5ex]
\hline
a & 16000 & 0 & - & - & 3780 & 7610\\
b & 2000 & 0 & - & - & 4000 & 7610\\
c & 2000 & 0 & - & - & 3510 & 9450\\
b & 16000 & 10 & 5910 & 0.104 & 2090 & 2700\\
c & 2000 & 10 & 3361 & 0.155 & 2870 & 4740\\
d & 2000 & 20 & 1028 & 2.39 & 800 & 800\\
e & 16000 & 30 & 1757 & 0.265 & 1890 & 1890\\
f & 2000 & 30 & 1148 & 0.0964 & 1830 & 9450\\
g & 2000 & 40 & 362 & 0.188 & 390 & 1190\\
h & 2000 & 50 & 247 & 2.95 & 170 & 300\\
i & 2000 & 50 & 1665 & 0.172 & 1340 & 2370\\
j & 2000 & 60 & 1260 & 0.415 & 2140 & 4740\\
k & 2000 & 70 & 1710 & 0.461 & 720 & 9450\\
l & 2000 & 80 & 1439 & 0.0751 & 1650 & 9450\\
m & 2000 & 80 & 518 & 0.281 & 1290 & 4740\\
n & 2000 & 90 & 381 & 1.81 & 210 & 2370\\
o & 16000 & 100 & 177 & 1.24 & 100 & 270\\
p & 2000 & 100 & 537 & 0.182 & 390 & 600\\
q & 2000 & 100 & 446 & 3.54 & 1630 & 4540\\
\hline
\end{tabular}
\caption{Major parameters of each simulation.  Columns in order are: (1) simulation name, (2) number of test particles, (3) star number density of early cluster environment, (4) minimum heliocentric distance of closest cluster star passage, (5) mass of closest passing cluster star, (6) minimum semimajor axis found in Oort Cloud at $t=4.5$ Gyrs, (7) isotropization semimajor axis in Oort Cloud at $t=4.5$ Gyrs.}
\label{table:2}
\end{table}

\begin{figure}[htp]
\centering
\includegraphics{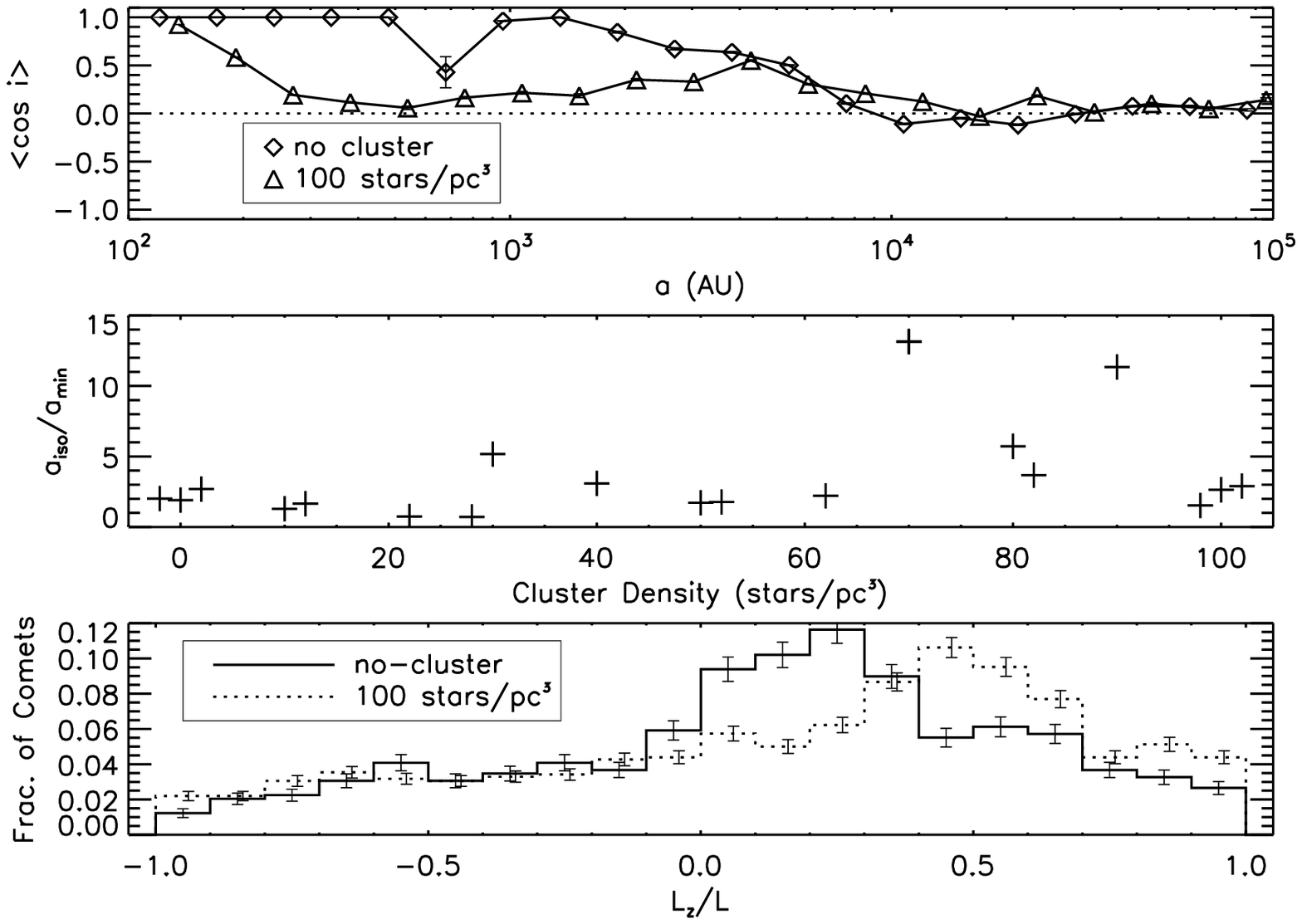}
\caption{\it{Top: }\rm{Plot of the mean cosine of cometary inclination with respect to the ecliptic vs. semimajor axis in the Oort Cloud of two simulations: the large no-cluster case (diamonds) and the large 100 stars/pc$^3$ cluster run (triangles).  }\it{Middle: }\rm{Ratio of the isotropization semimajor axis to the minimum semimajor axis in each Oort Cloud simulation at $t=4.5$ Gyrs vs initial cluster density.  Data are slightly offset to avoid confusion. }\it{Bottom: }\rm{}Distribution of azimuthal angular momentum (in galactic frame with respect to the plane of the galaxy) for comets in a simulation with no cluster environment (solid) and a simulation begun in a cluster density of 100 stars/pc$^{3}$ (dashed). Both of these distributions are taken at $t = 4.5$ Gyrs.}\label{fig:6}
\end{figure}

\subsection{Comet Trapping Efficiency}

We can also determine if the Sun's initial cluster environment has an impact on how efficiently comets are transferred to the Oort Cloud.  It has been proposed previously that immersion in an early cluster environment will increase the Oort Cloud trapping efficiency \citep{fern97}.  While this is true, the first two panels of Figure 7 show that the gain in the number of Oort Cloud objects is minimal.  This population increase is limited because the mechanism that increases the Oort Cloud capture rate also increases the Oort Cloud removal rate. Although the strong stellar perturbations efficiently shift comet perihelia outside of the planetary region, they also push the perihelia of existing Oort Cloud objects back into the planetary region where they will most likely be ejected.  Additionally, close encounters with cluster stars will put a large velocity impulse on the Sun relative to its Oort Cloud comets, which will strip the most tenuously held outer Oort Cloud layers to interstellar space.  

The gains from our cluster envirnoment are minimized because the rate that objects are being fed into the Oort Cloud via the scattered disk falls off rapidly as the population of objects near Jupiter and Saturn are depleted quickly at the beginning of the simulation.  In contrast, the stregth of stellar perturbations from cluster stars falls off much more slowly.  The number of objects in the planetary region orbiting inside 15 AU drops by $\sim$85\% within only 5 Myrs, whereas we immerse the solar system in a cluster environment for 100 Myrs.  It is possible that higher trapping efficiencies may be attained if the Sun were to have left the cluster earlier.  The first panel in Figure 7 demonstrates that this is indeed the case for each of our 3 large cluster simulations.  The Oort Cloud populations in each of the simulations peaks inside 25 Myrs.  Beyond this point, the cluster perturbations steadily erode the Oort Cloud population.  In addition, we see that for our densest cluster case, nearly 10\% of the particles are trapped in the Oort Cloud at one point before they are subsequently stripped.  

In the fourth panel of Figure 7, we plot the maximum trapping efficiency for all of our simulations during the first 100 Myrs.  As can be seen, trapping efficiencies as high as $\sim$12\% can be attained if the Sun exits a cluster at the right time.  In addition, in the third panel we plot the time at which the peak trapping efficiency was achieved for each simulation.  With the exception of one of our 100 stars/pc$^3$ simulations, all of the Oort Cloud populations peak within the first 30 Myrs.  In the one anamolous simulation, the Oort Cloud population remains roughly constant around 6\%, and a population peak is not well-defined.  We can conclude from all of this that had the solar system remained in a cluster for 100 Myrs or longer, a large fraction of the extra comets gained by the early cluster perturbations would be removed before the Sun transitioned to a field environment.

\begin{figure}[htp]
\centering
\includegraphics{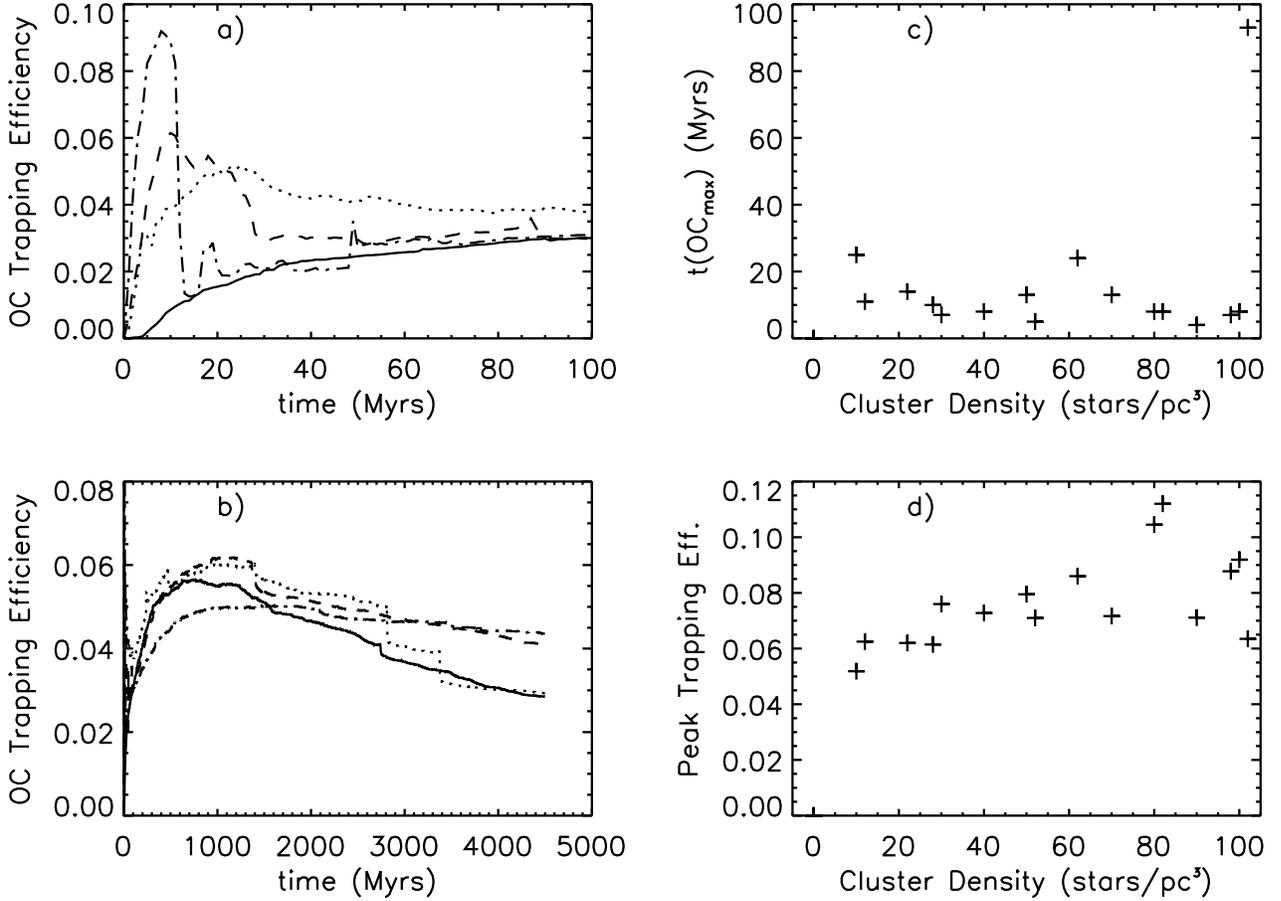}
\caption{\it{a: }\rm{Plot of trapping efficiency vs. time for the first 100 Myrs of our four large simulations. Simulations are as follows: }\it{Solid: }\rm{No-cluster, }\it{Dotted: }\rm{10 stars/pc$^3$ cluster, }\it{Dashed: }\rm{30 stars/pc$^3$ cluster, }\it{Dash-Dot: }\rm{100 stars/pc$^3$ cluster. }\it{b: }\rm{Plot of trapping efficiency vs. time for the entire time of our four large simulations. Simulations are as follows: }\it{Solid: }\rm{No-cluster, }\it{Dotted: }\rm{10 stars/pc$^3$ cluster, }\it{Dashed: }\rm{30 stars/pc$^3$ cluster, }\it{Dash-Dot: }\rm{100 stars/pc$^3$ cluster. }\it{c: } \rm{Time at which the maximum trapping efficiency is attained during the cluster phase for all of our cluster simulations vs. initial cluster density.  Data are slightly offset to avoid confusion. }\it{d: }\rm{Peak trapping efficiency attained during the cluster phase of all of our cluster simulations vs. initial cluster density.}}\label{fig:7}
\end{figure}

It has been suggested previously that an early open cluster environment will increase the relative contribution of Jupiter-Saturn zone comets to the Oort Cloud \citep{fern97}.  Previous simulations \citep{dun87,dones04} have shown that in the Sun's current galactic environment, the energy kicks of Jupiter and Saturn are so strong that most comets being scattered out by these planets are ejected to interstellar space before being trapped in the Oort Cloud.  Because orbital perihelia are torqued at smaller semimajor axes in clusters, there is a larger energy window into which the gas giants can scatter comets, and a greater fraction of comets can be trapped before they are ejected.  

To study this effect, we have plotted Oort Cloud trapping efficiency as a function of initial orbital perihelion in the planetary disk at the beginning of the simulation in Figure 8.  When doing this we have split our Oort Cloud into two populations of comets: those added during the cluster phase ($t < 100$ Myrs), which are shown in the top panel, and those added after the cluster phase ($t > 100$ Myrs), which is in the bottom panel.  These plots show that the gain of Jupiter-Saturn comets due to a cluster environment is minimal.  In the control case with no cluster environment, we see that less than 1\% of particles originating between 4 and 10 AU are trapped in the Oort Cloud.  In the denser cluster cases, this figure rises to roughly 2\%.  Although a cluster phase allows many more comets from the Jupiter-Saturn region to reach the Oort Cloud, most of these are quickly stripped by the same powerful stellar perturbations that make their trapping possible.  This effect was illustrated well in Figure 7a.  As stated previously, presumably a larger fraction of these comets could be stored in the Oort Cloud if the cluster dispersed earlier.

\begin{figure}[htp]
\centering
\includegraphics{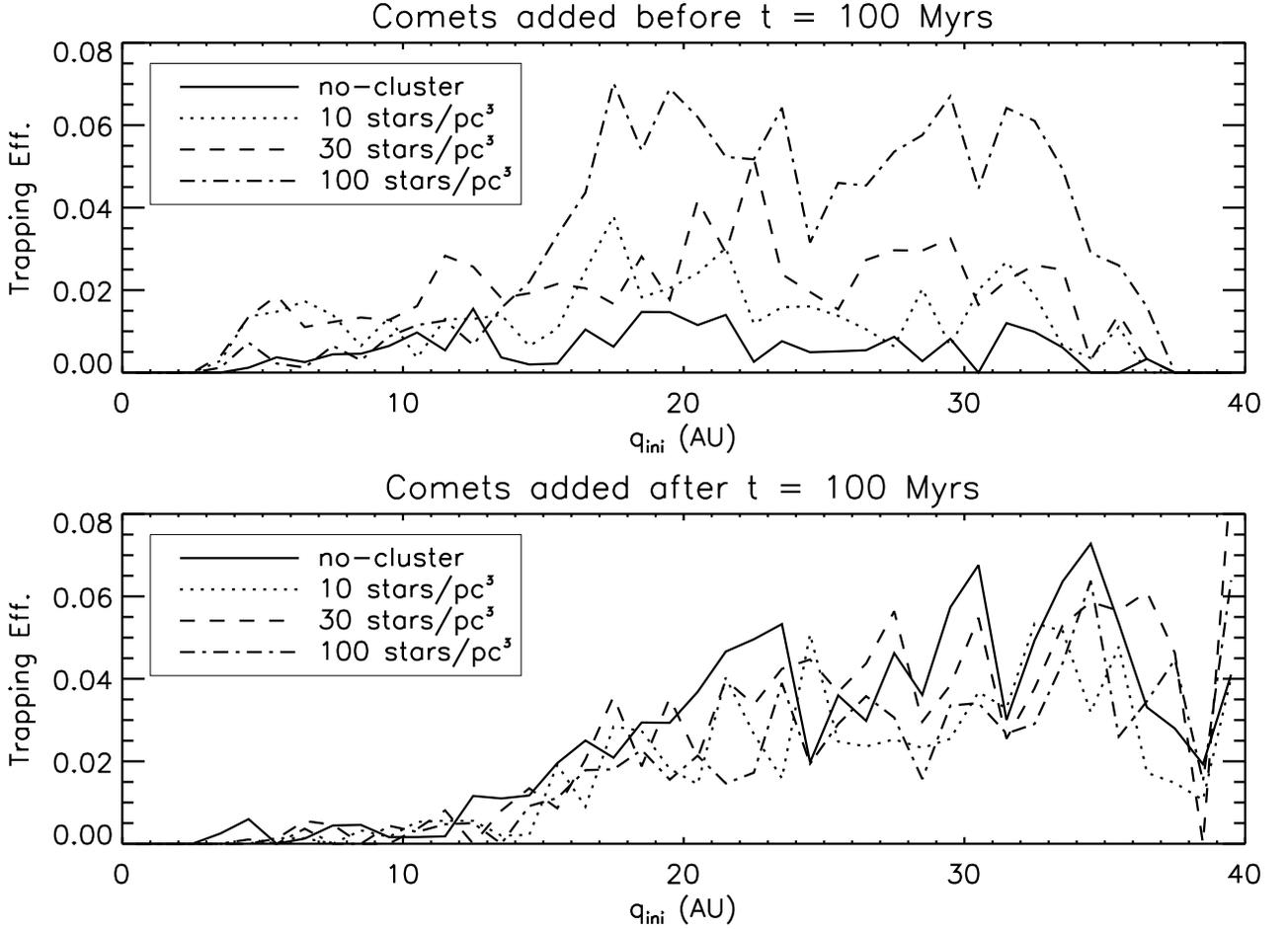}
\caption{\it{Top: }\rm{Plot of the trapping efficiency as a function of initial particle perihelion in the planetary disk for comets found in the Oort Cloud after 4.5 Gyrs but that entered the Oort Cloud during the first 100 Myrs of the simulation.  The results from our four large simulations are shown: }\it{Solid: }\rm{No-cluster, }\it{Dotted: }\rm{10 stars/pc$^3$ cluster, }\it{Dashed: }\rm{30 stars/pc$^3$ cluster, }\it{Dash-Dot: }\rm{100 stars/pc$^3$ cluster. }\it{Bottom: }\rm{Plot of the trapping efficiency as a function of initial particle perihelion in the planetary disk for comets found in the Oort Cloud after 4.5 Gyrs but that entered the Oort Cloud after the first 100 Myrs of the simulation.  The results from our four large simulations are shown: }\it{Solid: }\rm{No-cluster, }\it{Dotted: }\rm{10 stars/pc$^3$ cluster, }\it{Dashed: }\rm{30 stars/pc$^3$ cluster, }\it{Dash-Dot: }\rm{100 stars/pc$^3$ cluster. }}\label{fig:8}
\end{figure}

The second effect of the cluster phase seen in Figure 8 involves the trapping efficiencies of particles originating beyond 15 AU.  For these particles, a much higher fraction ($\sim$6\%) are trapped during the densest cluster environment, while only 1\% are trapped in the first 100 Myrs of the control simulation.  This is a result of the larger Oort Cloud energy window in a dense cluster.  Because the random walk in orbital energy takes place at a slower rate for particles controlled by Uranus and Neptune, very few can reach the classical Oort Cloud in the first 100 Myrs.  In contrast, the strong perturbations of the dense cluster environment reach much deeper into this outwardly diffusing scattered disk of particles.  During the time after the cluster phase, this effect is offset some because particles in the control case begin reaching the classical Oort Cloud in larger numbers, whereas trapping efficiencies are lower in the densest cluster case because the scattered disk has been somewhat depleted by the early strong perturbations.  

When examining Figure 8 it should be noted that an initial perihelion beyond 15 AU does not guarantee a comet will remain under the control of Uranus and Neptune.  In fact, most of these comets eventually migrate inward to Jupiter and Saturn before being scattered outward \citep{fern97,brasdun08}.  This process, however, happens over several tens of millions of years typically.  In the case of these inwardly migrating particles, once they reach Jupiter and Saturn, they are then scattered to large semimajor axis, and a small fraction reach the classical Oort Cloud in the control case.  In the cluster case, however, most of these are deposited in the inner Oort Cloud as long as the Sun remains in a cluster environment.  As with the depletion of the scattered disk discussed above, this effect also robs the outer Oort Cloud of potential bodies.  

The idea that the scattered disk could be depleted somewhat by early cluster perturbations is intriguing because producing a large disparity between the scattered disk mass and the outer Oort Cloud has been a problem for past models of Oort Cloud formation \citep{dones04}.  Past simulations of Oort Cloud formation have produced an outer Oort Cloud that is roughly 10 times the mass of the scattered disk.  In contrast, observations of long-period comets and Jupiter-family comets imply that the outer Oort Cloud should be more populous by a factor of 100-1000 \citet{dunlev97}.  Unfortunately, our simulations of an early open cluster environment do not alleviate this problem.  This is illustrated in the top panel of Figure 9 where the ratio of outer Oort Cloud population to scattered disk population is plotted for all of our simulations.  Here we classify an object as belonging to the scattered disk if $a > 45$ AU and $q < 45$ AU.  If anything, the addition of an open cluster environment seems to make this problem worse.  The highest ratio of outer Oort Cloud to scattered disk population is achieved in one of our no-cluster runs, whereas a few of our dense cluster cases actually have a ratio less than one.  

\begin{figure}[htp]
\centering
\includegraphics{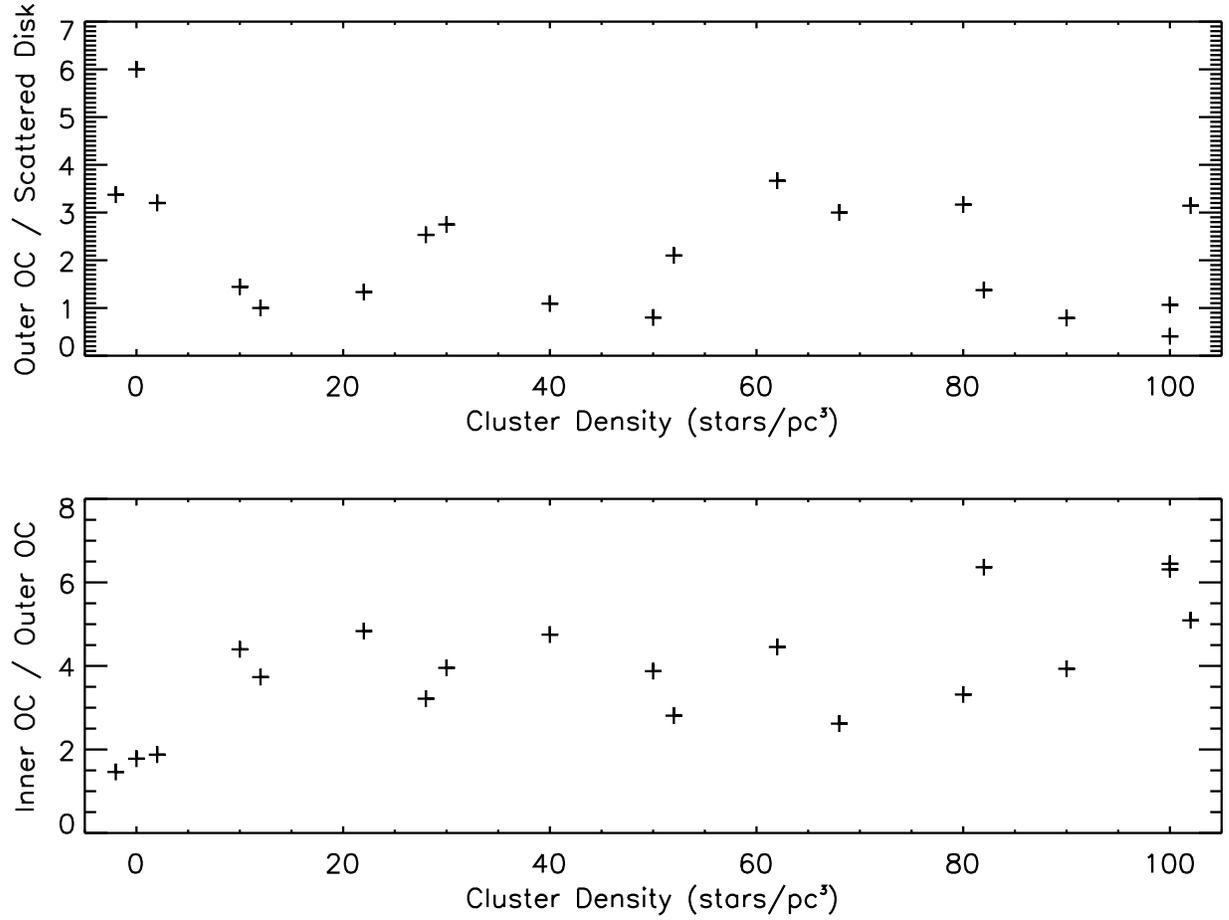}
\caption{\it{Top: }\rm{Population ratio of the outer Oort Cloud to scattered disk at $t=4.5$ Gyrs vs. initial cluster density for all of our simulations.  Data are slightly offset to avoid confusion. }\it{Bottom: }\rm{Population ratio of the inner Oort Cloud to the outer Oort Cloud at $t=4.5$ Gyrs vs. initial cluster density for all of our simulations.  Data are slightly offset to avoid confusion. }}\label{fig:9}
\end{figure}

We also look at the size of the inner Oort Clouds ($a <$ 20000 AU) for each one of our simulations in the bottom panel of Figure 9.  Here we plot the ratio of the inner Oort Cloud population to that of the outer Oort Cloud.  As expected from our previous analysis, it is clear that immersion in an open cluster environment will lead to a substantially more massive inner Oort Cloud relative to the outer one, although due to the stochasticity of this environment, there is not a tight correlation with cluster density.  \citet{dones04} find a size ratio of these two regions of $\sim$1, and our no-cluster runs find similar values between 1 and 2.  For some of our cluster simulations, however, we generate inner Oort Clouds that are more than a factor of 5 larger than the outer Oort Cloud, which more closely matches the results of \citet{dun87}.

This increase in inner Oort Cloud mass is due to the inward extension of the Oort Cloud energy window as discussed earlier.  This has two effects.  First, it allows many more comets to be captured into lower energy orbits than would be otherwise.  Secondly, this capture of comets into tightly bound orbits robs the outer Oort Cloud of some of the comets that would have otherwise ended up there.  

These two effects both act to lower the outer Oort Cloud trapping efficiency somewhat, and this highlights a second weakness of this and previous models of Oort Cloud formation.  The simulations discussed in \citet{dones04} find that only 2.5\% of their original planetesimals are found in the outer Oort Cloud after 4.0 Gyrs.  This value poses a problem for planet formation in the outer solar system, as it implies an original mass of $\sim150-300$ M$_{\Earth}$ to produce an outer Oort Cloud populous enough to be consistent with long-period comet observations \citep{dones04,weiss96}.  This value is 3-6 times more massive than the minimum-mass solar nebula.  This is such a high value that the solar system's actual configuration of planets would have been altered by either the formation of additional planets or extreme planet migration \citep{hahnmal99,thom02,gome04,tsig05}.  

We perform the same outer Oort Cloud population measurement for our simulations in the top panel of Figure 10.  We find even lower outer Oort Cloud trapping efficiencies than \citet{dones04}.  For our no-cluster cases, we find that only between 1 and 1.5\% of the original planetesimals are found in the Oort Cloud, which is only about half of the previously mentioned value.  It is not clear why this discrepancy exists.  We have slightly different initial conditions with the eccentricities and inclinations of our particles only going up to 0.01 and 0.02 radians, respectively, compared to their values of 0.02 and 0.01, but it seems unlikely this would make such a large difference.  Perhaps the low trapping efficiency problem is even worse than originally assessed in \citet{dones04}.  

A possible solution to this dilema, however, is that the real population of the outer Oort Cloud has been overestimated.  Using long-period comets discovered by the LINEAR survey, \citet{fran05} and \citet{nes07} have both concluded that previous estimates of the number of objects in the outer Oort Cloud were too high by at least an order of magnitude.  If this is indeed the case, then the requirements on both the solar nebula mass and the scattered disk implied from our simulations are less extreme.  Using the outer Oort Cloud population of 5 x 10$^{10}$ calculated in \citet{nes07} and assuming a mean comet mass of 4 x 10$^{16}$ g \citep{dones04}, we find that our simulations require only a $20-65$ M$_{\Earth}$ initial disk of planetesimals.  

We also see in the top panel of Figure 10 that some of our cluster simulations yield even lower outer Oort Cloud populations, with the lowest just above 0.5\%.  However, others have outer Oort Clouds as large as the no-cluster cases.  It appears that immersion in a cluster environment does not necessarily lead to a more anemic outer Oort Cloud.  This seems to imply that the majority of comets in the outer Oort Cloud originate in the Uranus-Neptune region since the migration toward Jupiter and Saturn and subsequent scattering of bodies near these planets proceeds on a timescale of hundreds of Myrs.  Indeed, the plot of trapping efficiency as a function of initial distance in the planetary disk seems to indicate this as well.  

To more directly compare the effect of our cluster environments with those modeled in \citet{bras06}, we additionally plot the total Oort Cloud trapping efficiency immediately after the cluster phase has ended ($t=$100 Myrs) in the bottom panel of Figure 10.  In this figure, we see that our trapping efficiencies range from 1\% up to 7\%.  This is less than the efficiencies of 2-18\% found in \citet{bras06}.  This comparison is not entirely accurate, however, because \citet{bras06} only plot the fraction of ``dynamically active'' comets ($e > $0.1 or $q > $35 AU) that are trapped.  Using the same criteria on our simulations raises our trapping rates by about a factor of 1.25 for each simulation.  Again due to the random nature of the most powerful cluster perturbations, there is no clear correlation between cluster density and trapping efficiency.  

\begin{figure}[htp]
\centering
\includegraphics{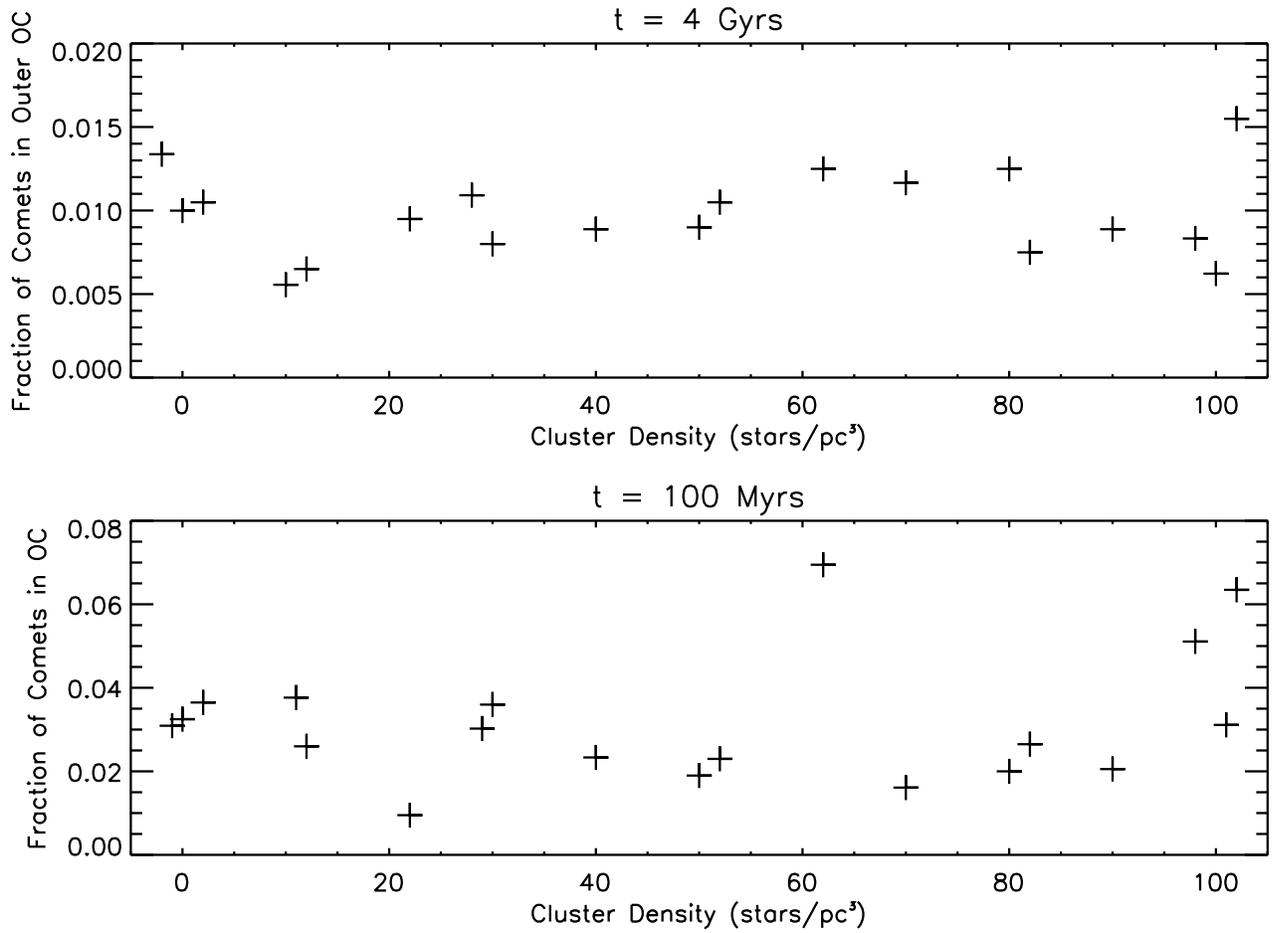}
\caption{\it{Top: } \rm{Fraction of all comets that are found in the outer Oort Cloud after 4 Gyrs vs. initial cluster density.  Data are offset to avoid confusion. }\it{Bottom: }\rm{Fraction of all comets that are found in the whole Oort Cloud after 100 Myrs vs. initial cluster density.  Data are offset to avoid confusion. }}\label{fig:10}
\end{figure}

\subsection{Diffusion of the Inner Oort Cloud}

As already discussed, one of the weaknesses of simulations of Oort Cloud formation is the small fraction of planetesimals that wind up in the Oort Cloud.  \citet{weiss96} suggested that this problem could  possibly be solved if the outer portions of the Oort Cloud were steadily replenished over time by the outward diffusion of a massive inner Oort Cloud.  This scenario is advantageous because it provides a second dynamical pathway for bodies to wind up in the outer Oort Cloud.  Rather than direct placement into the outer Oort Cloud from the scattered disk (the bulk of which takes place in the first Gyr), bodies can be stored for long periods in the inner Oort Cloud before they dynamically diffuse outward to bolster the outer Oort Cloud population as it is eroded by constant perturbations.  

We look for this effect in our simulations by recording the semimajor axis each comet has when it first reaches the Oort Cloud ($q >$ 45 AU).  At the end of our simulations we then compare each comet's final semimajor axis with its initial one.  Our results of this analysis for the outer Oort Cloud ($a >$ 20000 AU) in each of our four large simulations is shown in Figure 11.  In this figure, we have plotted the fraction of outer Oort Cloud comets that entered the Oort Cloud originally inside a given semimajor axis.  This distribution of initial semimajor axes looks very similar for three of our four large simulations.  In these simulations, roughly half of the outer Oort Cloud comets have always had semimajor axes greater than 20000 AU, and only 10-15\% diffused outward from initial semimajor axes smaller than 10000 AU.  The contribution to the outer Oort Cloud from comets diffusing from the inner 5000 AU is under 10\% in each of the simulations.  The semimajor axis distribution for the large 10 stars/pc$^3$ simulation looks markedly different from the other three.  This is no doubt a result of the numerous close field star encounters that the solar system suffered in this simulation.  As stated previously, this is a highly improbable occurance, and this particular run should not be considered representative of typical Oort Cloud evolution.  

\begin{figure}[htp]
\centering
\includegraphics{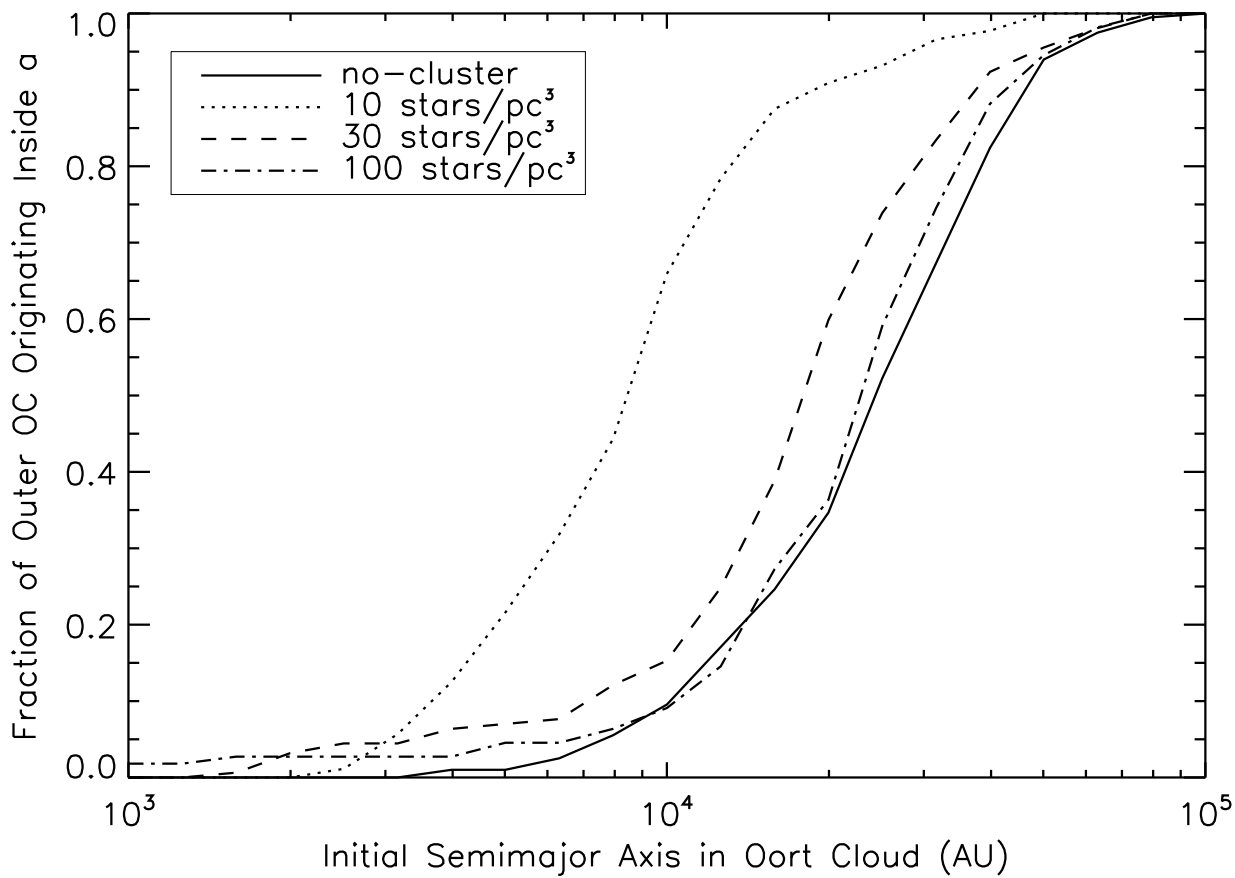}
\caption{Cumulative distribution function of the initial Oort Cloud semimajor axes of all the comets in the outer Oort Cloud after 4.5 Gyrs in our four large simulations.  The simulation data are as follows: \it{Solid: }\rm{No-cluster, }\it{Dotted: }\rm{10 stars/pc$^3$ cluster, }\it{Dashed: }\rm{30 stars/pc$^3$ cluster, }\it{Dash-Dot: }\rm{100 stars/pc$^3$ cluster. }}\label{fig:11}
\end{figure}

We can conclude from Figure 11 that the degree of outer Oort Cloud population bolstering done by the inner Oort Cloud is fairly limited in semimajor axis range.  There appears to be significant outward migration of comets in the 10000-20000 AU, but inside this region diffusion to the outer Oort Cloud is negligible.  Although an open cluster environment does significantly enhance the amount of material in the inner Oort Cloud, we have seen in Figures 2 and 3 that the bulk of this material is placed in the inner few thousand AU of the Oort Cloud.  Because of this, most of these comets will remain locked in the extreme inner region of the Oort Cloud, and the gains in the inner Oort Cloud population due to the Sun's early environment will not be reflected in the rate of diffusion to the outer Oort Cloud.  

\subsection{Replication of the Extended Scattered Disk}

Our modeling work as well as previous simulations of slow, close stellar passages \citep{lev04} illustrate that the perturbations of an open cluster environment penetrate much deeper into the scattered disk compared to normal field perturbations.  \citet{morblev04} posited that these types of encounters may have been reponsible for producing the orbits of extended scattered disk bodies like Sedna, 2000 CR$_{105}$, 2003 UB$_{313}$, and Buffy (2004 XR$_{190}$).  All of these orbits are only produced by significant perturbations, yet their semimajor axes are all hundreds of AU or less, effectively insulating them from the external perturbations the Sun currently experiences.  An early star cluster environment is a plausible setting where stellar encounters necessary to perturb orbits this small would be common.  

Indeed, \citet{bras06} demonstrate that numerous embedded cluster environments are capable of generating the orbits of real extended scattered disk objects.  To determine if similar results can be achieved with an open cluster environment, we replicate their analysis by defining ranges in the orbital elements of our particles that correspond to the orbits of real extended scattered disk objects.  These ranges that correspond to each object are listed in Table 3.  In Figure 12, we plot the percentage of combined scattered disk and Oort Cloud particles that are found on orbits similar to Sedna, 2000 CR$_{105}$, and 2003 UB$_{313}$ at the end of each of our simulations.  A plot for Buffy is not shown, as we were unable to generate a similar orbit in any of our simulations.  As can be seen in the figure, 2003 UB$_{313}$ is the most difficult object to produce in our simulations.  Only one simulation, the large 100 stars/pc$^3$ one, produced an object like UB$_{313}$.  In the case of 2000 CR$_{105}$, four different simulations produced orbits similar to this object.  Lastly, orbits resembling Sedna's were produced in five of our 16 cluster simulations.  The obvious correlation between cluster density and the prevalence of these orbits that is seen in \citet{bras06} is absent in our results.  Once again, this is a consequence of the stochatistic nature of the strongest open cluster perturbations.  

\begin{table}[htp]
\centering
\begin{tabular}{c c c c}
\hline
Object & Range in \it{a} \rm{(AU)} & Range in \it{q} \rm{(AU)} & Range in \it{i} \rm{($^\circ$)}\\ [0.5ex]
\hline
Buffy & 50-65 & 45-60 & 40-180\\
2003 UB$_{313}$ & 53-80 & 37-46 & 37-52\\
2000 CR$_{105}$ & 200-300 & 40-50 & 15-30 \\
Sedna & 400-600 & 68-84 & 0-180\\
\hline
\end{tabular}
\caption{Range of particle orbital elements corresponding to real objects in the extended scattered disk.}
\label{table:3}
\end{table}

\begin{figure}[htp]
\centering
\includegraphics{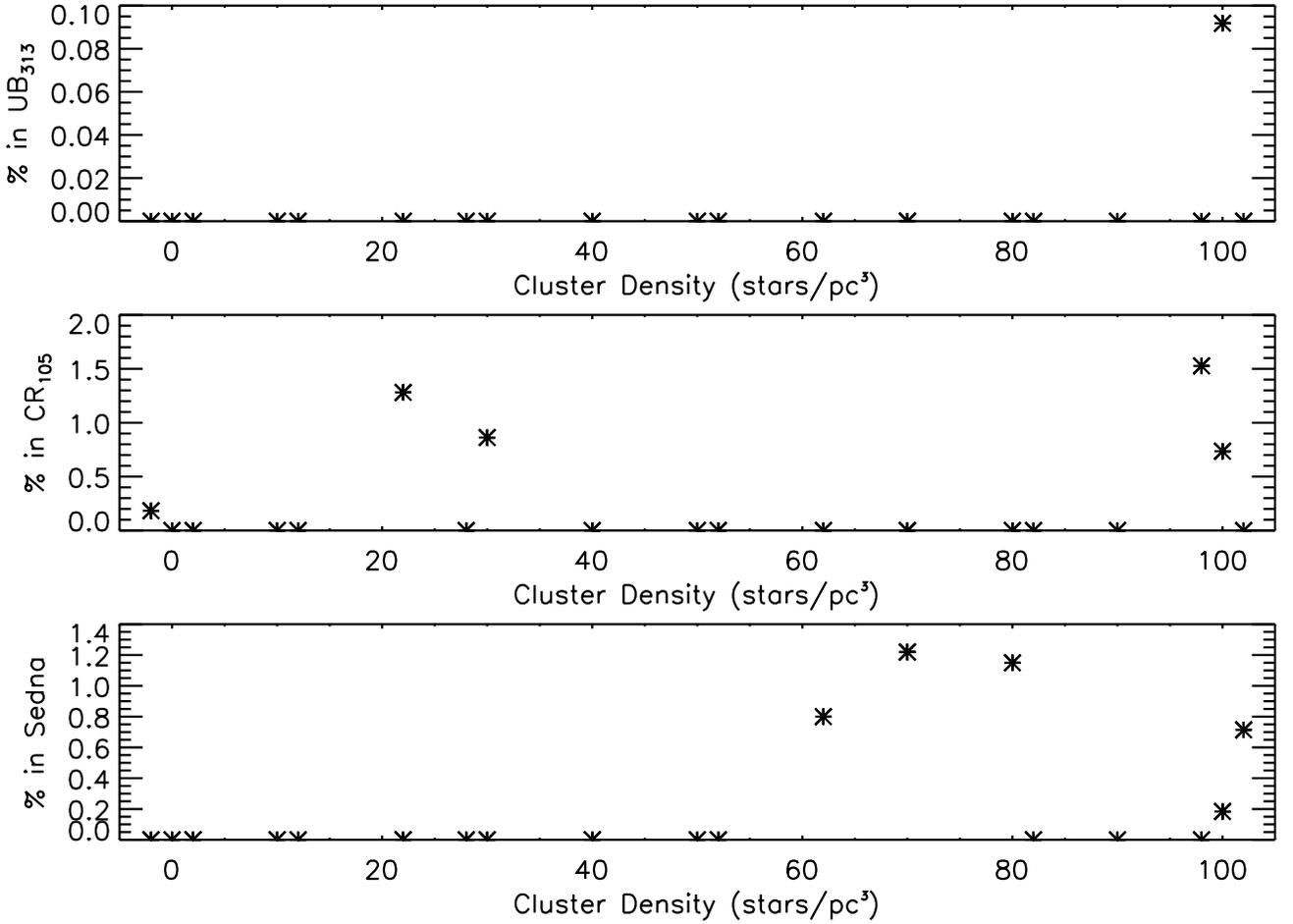}
\caption{Fraction of comets found on 2003 UB$_{313}$-like orbits (\it{top-panel}\rm{), 2000 CR$_{105}$-like orbits (}\it{middle-panel}\rm{), and Sedna-like orbits (}\it{bottom-panel}\rm{). Data are slightly offset to avoid confusion.}}\label{fig:12}
\end{figure}

It should be noted that only one of our simulations generates the orbits of Sedna, 2000 CR$_{105}$, and 2003 UB$_{313}$ simultaneously as defined by \citet{bras06}.  This is the large 100 stars/pc$^3$ simulation.  In panels 13a and 13b, plots of the distribution of orbital elements for all of the particles in this simulation with $a <$ 1000 AU demonstrates that the range of orbital elements for all three of these objects is indeed sampled by our simulation.  For comparison, we have also made similar plots for the large 30 stars/pc$^3$ simulation in panels 13c and 13d.  We have decided not to include similar plots for our large 10 stars/pc$^3$ simulation because the typical orbital element distribution has been disrupted by numerous close field star passages.  Given the orbital element distributions seen in 13a and 13b, we feel that it is in fact possible to have a single open cluster environment give rise to each of these three extended scattered disk orbits.  

\citet{bras06} point out that each of their simulations that generates an analogue to UB$_{313}$ also experiences such strong cluster perturbations that their Kuiper Belt is stirred inside 20 AU, i.e. the mean eccentricity of planetesimals exceeds 0.05.  When this occurs, collisions between individual planetesimals tend to be erosive rather than accretionary \citep{kenbrom02}.  Because it is generally thought that the primordial Kuiper Belt extended out to a minimum of 30 AU \citep{gome04}, \citet{bras06} conclude that the orbit of 2003 UB$_{313}$ is not the result of embedded cluster perturbations.  

Our simulation that produces a UB$_{313}$ analogue also contains 625 particles with semimajor axes between 33 and 40 AU after 4.5 Gyrs.  Dividing these particle orbits into 1 AU bins and taking it as a Kuiper Belt analogue, we have checked for Kuiper Belt stirring as well.  In contrast to the results of \citet{bras06}, we find mean eccentricities between 0.02 and 0.04 for each of the bins inside 40 AU.  It seems then that at least one open cluster scenario can reproduce the orbits of 2003 UB$_{313}$, Sedna, and 2000 CR$_{105}$ without causing collisional erosion in the inner primordial Kuiper Belt.  As stated previously, however, no simulation is capable of forming a Buffy analogue and a different mechanism is most likely required to produce this object's orbit.

Another issue that arises when invoking external perturbations to form the extended scattered disk concerns the prevalance of objects with perihelion between 40 and 50 AU.  \citet{lev04} model the effect that single close stellar passages have on the scattered disk and Oort Cloud.  In their work, they find that close stellar encounters consistently produce a scattered disk that has comparable numbers of bodies with 30 $< q <$ 40 AU and 40 $< q <$ 50 AU.  When extrapolating a perihelion distribution of our real scattered disk, however, they determine that only $\sim$5\% of scattered disk objects are on orbits with 40 $< q <$ 50 AU.  We, therefore, look at the relative sizes of these populations in our large 100 stars/pc$^3$ simulation as well.  Like \citet{lev04}, we find that just under half (48\%) of the particles with 30 $< q <$ 50 AU have perihelia between 40 and 50 AU.  Thus, it appears this aspect of an open cluster environment which reproduces most of the known extended scattered disk objects is at odds with outer solar system observations.  

There is also one other unobserved consequence of an open cluster environment that forms the extended scattered disk.  This is a retrograde component of the scattered and extended scattered disks.  Unlike most of our runs, the large 100 stars/pc$^3$ simulation produces a significant population of retrograde orbits in both of these areas of the solar system.  This is illustrated well in Figure 14 where we have plotted all of the retrograde orbits with 0 $< q < $ 200 AU in each of our four large simulations.  In our densest cluster simulation, $\sim$2\% of the bodies with $a < 1000$ AU and $30 < q < 50$ AU are on retrograde orbits, whereas the other three simulations contains no retrograde counterparts in this regime.  Because this percentage is so small, determining whether a retrograde population exists will require many more detections of scattered and extended scattered disk bodies.  Note that the few bodies seen with $q < 30$ AU and small $a$ are rare returning Oort Cloud comets whose semimajor axes have been drawn back down through interactions with the planets.

If the formation of Sedna, 2000 CR$_{105}$, and 2004 UB$_{313}$, are due to the perturbations of an early cluster environment, then there should exist a smaller population of retrograde objects with similar perihelia and semimajor axes.  Because searches for objects in the Kuiper Belt and Scattered disk are centered on the ecliptic, we may have not yet discovered this possible population of retrograde bodies.  Other scenarios put forth to explain the orbits of the extended scattered disk all produce the observed high inclinations, but they do not produce any retrograde orbits \citep{gom05b,gladchan06}.  The presence or absence of this population would serve as a test for this particular formation process.  Currently, only very few extended scattered disk objects are known.  In the coming years, however, new sky surveys such as PanSTARRS and LSST will undoubtedly discover many new members of this population, and these surveys will not be biased toward the discovery of low inclination orbits.

If the extended scattered disk is indeed a dynamical footprint left by an early cluster environment, our simulations show that this has large implications for the structure of the Oort Cloud's interior.  Normalized by the number of comets in the outer Oort Cloud ($a >$ 20000 AU), we find that the population of the Oort Cloud's inner region can vary by nearly an order of magnitude depending on the Sun's early environment.  In this way, the production mechanism of the extended scattered disk is intimately linked to the abundance of the inner Oort Cloud.

\begin{figure}[htp]
\centering
\includegraphics{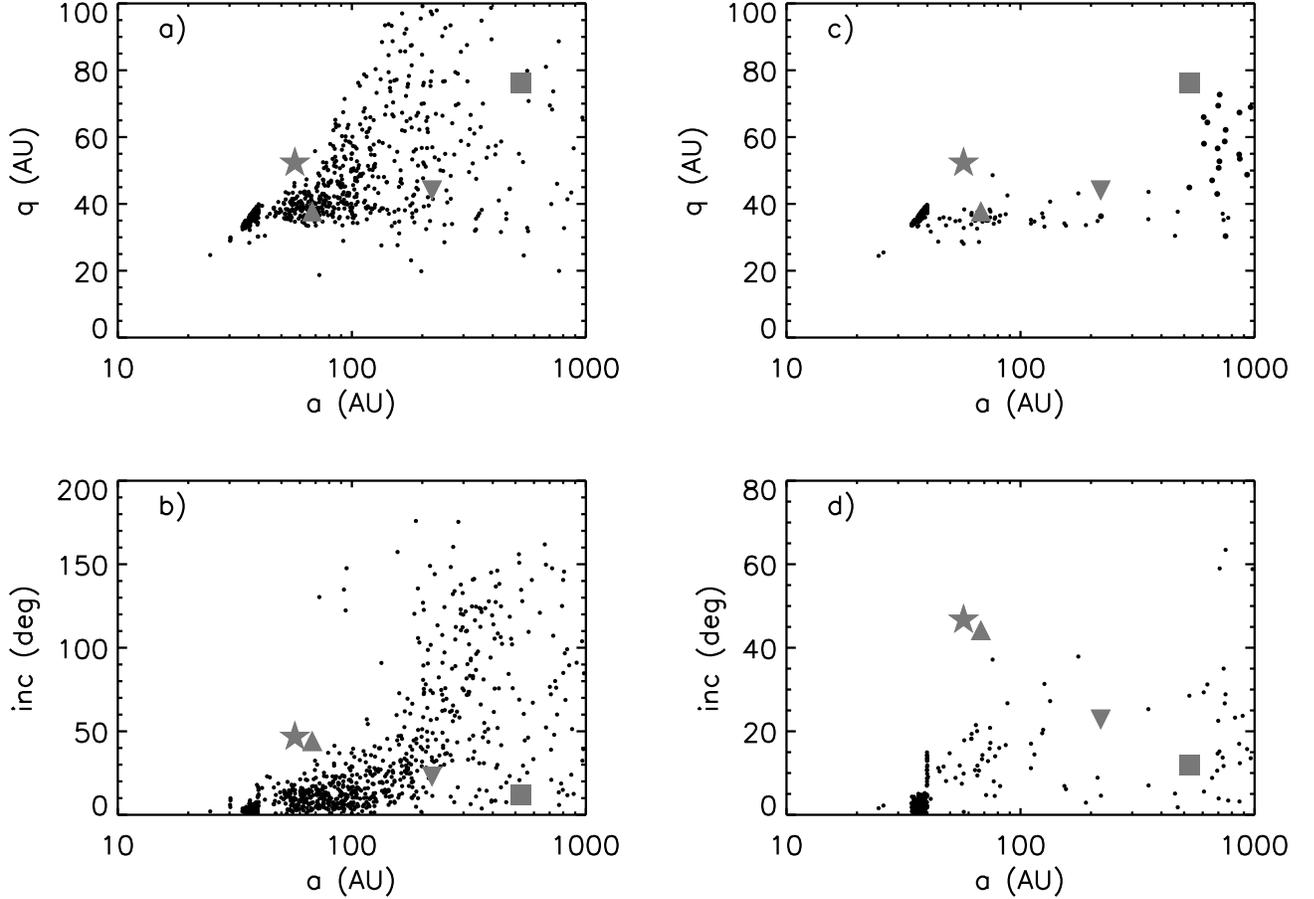}
\caption{\it{a: }\rm{Plot of perihelion vs. semimajor axis for scattered disk and Oort Cloud particles for our large 100 stars/pc$^3$ cluster simulation after 4.5 Gyrs. }\it{b: }\rm{Plot of inclination vs. semimajor axis for scattered disk and Oort Cloud particles for our large 100 stars/pc$^3$ cluster simulation after 4.5 Gyrs. }\it{c: }\rm{Plot of perihelion vs. semimajor axis for scattered disk and Oort Cloud particles for our large 30 stars/pc$^3$ cluster simulation after 4.5 Gyrs. }\it{d: }\rm{Plot of inclination vs. semimajor axis for scattered disk and Oort Cloud particles for our large 30 stars/pc$^3$ cluster simulation after 4.5 Gyrs. }\it{Note: } \rm{The orbits of Buffy (star), 2003 UB$_{313}$ (triangle), 2000 CR$_{105}$ (upside-down triangle), and Sedna (square) are plotted in each panel as well.}}\label{fig:14}
\end{figure}

\begin{figure}[htp]
\centering
\includegraphics{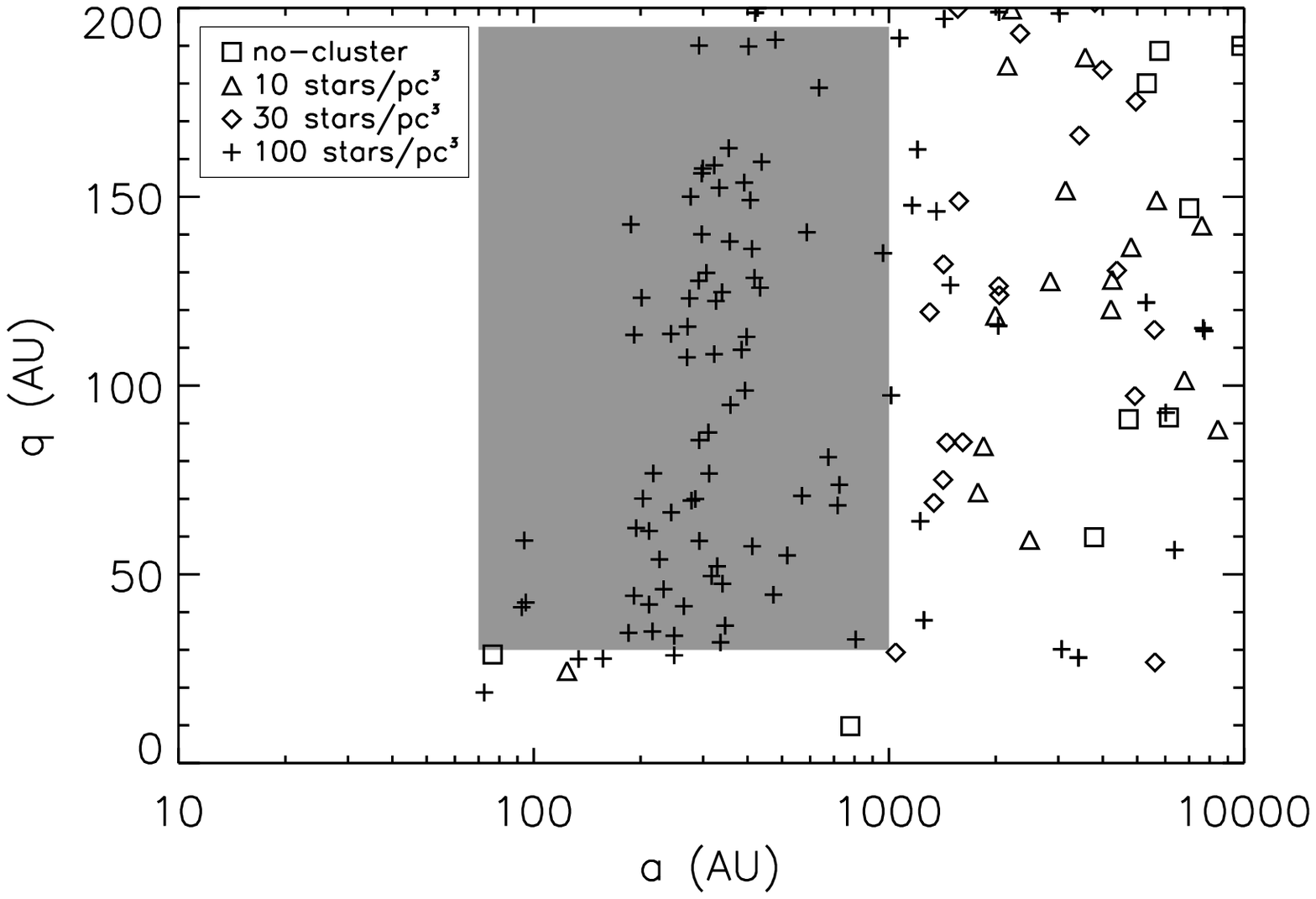}
\caption{Plot of the semimajor axis vs. perihelion for all of of the retrograde orbits in each of the four Oort Cloud simulations at $t=$ 4.5 Gyrs.  Cluster environment densities are as follows: \it{Squares:} \rm{No cluster environment,} \it{Triangles:} \rm{10 stars/pc$^{3}$,} \it{Diamonds:} \rm{30 stars/pc$^{3}$,} \it{Crosses:} \rm{100 stars/pc$^{3}$. The region of interest for the scattered disk and extended scattered disk is highlighted by the shaded area.}}\label{fig:13}
\end{figure}

\section{Summary}

By modeling the formation of the Oort Cloud in numerous different open cluster environments, we explore the effects early strong stellar perturbations have on this process.  Because scattered disk objects do not need to reach as large of a distance before they feel the effects of stellar encounters, these cluster environments preferentially enrich the inner Oort Cloud.  We determine that an open cluster environment can produce Oort Cloud objects with semimajor axes as small as $\sim100$ AU.  In addition, we find that the degree of Oort Cloud central concentration is determined largely by the closest encounter with a cluster star.  Because of this dependence on a single event in each simulation, Oort Cloud formation in an open cluster environment is a very stochastic process with only a weak dependence on the specific cluster density.  

While the stronger cluster perturbations pull material into the Oort Cloud much more efficiently, the variance in the whole cloud's population and demographics is small.  This is because these perturbations are also more efficient in removing existing bodies from the Oort Cloud.  The result is that the mean residence time of a comet in the Oort Cloud is drastically reduced while the Sun resides in a cluster.  Our work also shows, however, that if our cluster environments dispersed in tens of Myrs, 2-3 times more bodies, mostly those from the Jupiter-Saturn zone, can be trapped in the inner Oort Cloud.

With respect to only the outer Oort Cloud, the impact of an open cluster environment appears fairly insignificant.  This is mostly due to the fact that the outer cloud continues to form and dynamically evolve long after the cluster phase has ended in our simulations.  Examining our entire set of simulations, we see that there is no well-defined correlation between cluster environments and the population size of the outer Oort Cloud.  In addition, we find that in all of our simulations the outer Oort Cloud trapping efficiencies range between only 0.5\% and 1.5\%.  This seems to put an even more extreme mass requirement on the primordial solar nebula than previous works.  However, using the most recent outer Oort Cloud population estimates from LINEAR survey data  \citep{nes07}, our simulations only imply a 20-65 M$_{\Earth}$ initial disk of planetesimals.

Unlike the outer Oort Cloud, the compact inner cloud formed in the cluster environment has a very short period of dynamical evolution.  Once the solar system leaves its primordial environment, orbital evolution stops for nearly all of these comets.  The reason for this is that the external perturbations of a field environment are too weak to penentrate into the inner regions where these bodies orbit.  As a result, there is only limited outward diffusion of the inner Oort Cloud.  Our simulations show that the vast majority of outer Oort Cloud comets enter the Oort Cloud with orbital semimajor axes larger than 10000 AU.  Hence, the massive inner Oort Clouds produced by clusters are not very useful in bolstering the population of the steadily eroding outer cloud.  

Furthermore, the only time the inner Oort Cloud is significantly perturbed occurs during ``comet showers,'' which are triggered by rare extremely close stellar encounters as well as during penetrating encounters with GMCs.  The potency of these showers depends on the particular encounter as well as the population of the inner Oort Cloud, which we have shown is a function of the Sun's early environment.  This is an aspect with possible astrobiological consequences that we intend to explore in a forthcoming paper.

Regardless of the actual primordial environment of the Sun, our work shows it will play a large role in shaping the structure and abundance of today's inner Oort Cloud.  If the orbits of extended scattered disk objects are the result of a dense primordial environment, then the inner Oort Cloud must be closer and more populous than the predictions of previous simulations of Oort Cloud formation.  Our current sample of long-period comets sheds no light on this issue, however, as they all originate in the outer Oort Cloud.  Thus, observations of more distant objects are needed.  In particular, a sample of comets with perihelia outside Saturn's orbit will probe the entire Oort Cloud.  Additionally, expanding the catalogue of extended scattered disk bodies, including searches for bodies highly inclined to the ecliptic, will be invaluable in constraining the dynamical history of the solar system as well as the unobserved regions of the Oort Cloud.

\section{Acknowledgments}

We would like to thank the reviewers Ramon Brasser and Julio A. Fern\'{a}ndez for insightful comments and suggestions that greatly improved the quality of this work.  We would also like to thank the NSF (grant AST-0709191) and the NAI for funding the majority of this work.  Finally, the bulk of our computing work was performed using the Purdue Teragrid computing facilities managed with Condor scheduling software (see http://www.cs.wisc.edu/condor).

\bibliography{OCFormation}

\end{document}